\documentclass[review]{elsarticle}

\newif\ifbiorxiv

\biorxivtrue

\ifbiorxiv
    \usepackage{hyperref}
    \usepackage[a4paper, total={6in, 10in}]{geometry}
    \usepackage{setspace}
\else
    \usepackage{lineno,hyperref}
    \modulolinenumbers[5]
    \journal{Computational and Structural Biotechnology Journal}
    \linenumbers
\fi

\begin{document}

\begin{frontmatter}

\title{Computational challenges of cell cycle analysis using single cell transcriptomics}
\author{Alexander Chervov$^{1,2,3}$, Andrei Zinovyev$^{1,2,3}$}
\address{$^1$Institut Curie, PSL Research University, Paris, France \\
$^2$INSERM U900, Paris, France \\
$^3$MINES ParisTech, CBIO-Centre for Computational Biology, PSL Research University, Paris, France.
}

\begin{abstract}
The cell cycle is one of  the most fundamental biological processes important for understanding normal physiology and various pathologies such as cancer. 
Single cell RNA sequencing technologies give an opportunity  to analyse the cell cycle transcriptome dynamics
in an unprecedented range of conditions (cell types and perturbations), with thousands of publicly available datasets. 
Here we review the main computational tasks in such analysis:
1) identification of cell cycle phases, 
2) pseudotime inference,
3) identification and profiling of cell cycle-related genes, 
4) removing cell cycle effect,
5) identification and analysis of the G0 (quiescent) cells.
We review seventeen software packages that are available today for the cell cycle analysis using scRNA-seq data. 
Despite huge progress achieved, none of the packages 
can produce complete and reliable results with respect to all aforementioned tasks. One of the major difficulties for existing packages is distinguishing between two patterns of cell cycle transcriptomic dynamics:
normal and characteristic for embryonic stem cells (ESC), with the latter one shared by many cancer cell lines. Moreover, some cell lines are characterized by a mixture of two subpopulations, one following the standard and one ESC-like cell cycle, which makes the analysis even more challenging. In conclusion, we discuss the difficulties of the analysis of cell cycle-related single cell transcriptome and provide certain guidelines for the use of the existing methods.


\end{abstract}


\end{frontmatter}

\newpage
\tableofcontents

\section{Introduction}

\subsection{The context of the review}

The cell cycle is a fundamental biological process 
which is of utmost importance for cancer research, where the deviation from the normal cell cycle progression is expected as well as for other domains.
It is widely studied from various points of view: for example, the Nobel prize in 2001 was awarded for 
"discoveries of key regulators of the cell cycle”. It is still
a field of active research with more than 10000 publications per year containing the keyword "cell cycle" \cite{Patterson2021CDKCP}. 

Single cell sequencing technologies had a huge impact on many areas of the biological research in particular: oncology \cite{tirosh2019deciphering},
developmental biology \cite{griffiths2018developmental},  immunology \cite{neu2017single}, etc.
 
The aim of the present review is to discuss how the properties of cell cycle in a given context can be studied using the scRNA-seq
data. We briefly describe 17 available packages for such analysis, focusing on computational, rather than biological, aspects. We also focus on cell cycle for mammals because such data widely dominate available scRNA-seq datasets and for their potential impact on translational research.

{\bf Navigation through the review.} 
Readers of this review oriented towards application of computational methods might have a question on how  good are the most popular software packages for single cell data analysis (e.g., Seurat \cite{Butler2018seuratV2}, Scanpy \cite{wolf2018scanpy}) to analyse the cell cycle dynamics and are there better alternatives? The brief 
analysis of the Seurat/Scanpy features related to cell cycle can be found in section \ref{subsectCCphasesWhatPhases},
\ifbiorxiv
\ref{sectSeuratScanpy},
\fi
some our brief conclusions on the choice of the package might be found in section \ref{sectSummary}.
Readers with more biological focus may be first interested in the question:
what are the new insights can scRNA-seq methods provide compared to the more traditional  methods, when cell cycle is studied? This aspect is discussed in the subsections \ref{sect_cellcycle_scrnaseq}, \ref{sectAdvDisadvSCRNAseq}. 
Despite many limitations, unprecedented amount of publicly available scRNA-seq data 
provides a previously not available opportunity to analyse the cell cycle dynamics under 
various conditions (cell types, perturbations (drugs, knockouts, mutations)), using only computational approach. 

The organization of the paper is the following: 
subsection \ref{sectBriefOverview} provides brief snapshot of the review - the main computational tasks formulated, the packages listed,
the main subtle points are highlighted. Section \ref{sectBulkTranscriptomicStudies} - brief reminder of biological research related to the review;
section \ref{basicIdeas} - highlights  basic ideas and subtleties of the cell cycle analysis;
section \ref{sectTasks} - detailed discussion of the main tasks: descriptions, subtleties, approaches, state of the art, etc.;
section \ref{sectDiscussion} - summary, guidelines, open questions, concluding remarks
\ifbiorxiv
; section \ref{sectPackages} provides a technical review of 17 existing packages
\fi.
Sections \ref{basicIdeas}, \ref{sectTasks}  are somewhat key for the review.

The review is an outcome of developing our own methodology to analyse the cell cycle using single cell RNA sequencing data and comparing it with existing approaches. More details on this are at end of the section \ref{sectBriefOverview}.

{\bf Single cell RNA sequencing.}
The standard single cell RNA sequencing biotechnologies combined
with bioinformatics pipelines produce  "count matrices",
i.e. cell x gene (transcript) matrices which characterizes the level of expression of each gene (transcript) in every cell captured for the measurement.  The technology rapidly evolves since 2014, producing large amounts of data.  The so-called Svennson's list contains references to more than 1600 scRNA-seq experiments and it is not exhaustive, with the median number of cells in a dataset equal to 10000 cells \cite{Svensson2020}. Top experiments delivered more than 4 millions individual cell transcriptomes (e.g., the human fetal atlas \cite{Cao2020} ).
There were several atlas level studies published:
"Tabula Muris" with 100,000 mouse cells from 20 organs and tissues \cite{tabula2018single}, "Tabula Sapiens" with more than 500 000 cells from 400 cell types and 24 tissues and organs \cite{tabula2022Sapiens},
"ENCODE" with sequenced mouse embryos from day 10.5 to the birth, sampling 17 tissues and organs \cite{He2020}. The Human Cell Atlas Project is an international collaborative effort aiming at defining all human cell types in terms of distinctive molecular profiles \cite{Regev2017}. Several other atlases are available in the field, also many cancer cell lines are profiled: i.e., part of Cancer Cell Line Encyclopedia (CCLE) with 198 cancer cell lines from 22 cancer types \cite{Kinker2020}, 35,276 cells from 32 breast cancer cell lines \cite{Gambardella2022}.
Datasets for various cell types, cell lines, under various perturbations (drugs, knockouts, diseases ) are publicly available.  Recent breakthrough technology Perturb-seq  merges CRISPR with scRNA-seq ad provides data on perturbations systematically and on unprecedented scale \cite{REPLOGLE2022}, \cite{Ursu2022}. 



{\bf The cell cycle.} The cell cycle, or cell-division cycle, is a biological process starting from
the cell birth to its division into two daughter cells that give birth to two new cells that in turn grow and divide into two daughter cells, and so on. Therefore, it is a cyclic process. 
The eukaryotic cell cycle is known to be subdivided into four main phases:
G1 (Gap 1), S (synthesis - replicating the DNA), 
G2 (Gap 2), M (Mitosis - cell division).
It is highly conservative for eukaryotes (but is somewhat different from bacteria). For example, the Nobel prize winning studies were mainly done for the yeast, but similar genes and mechanisms are present in mammalian cells despite billions of years of evolutionary distance. 
The duration of the cell cycle for normal mammalian cells varies around 24 hours, while 
for some faster dividing mammalian cells (like embryonic stem cells (ESC)) it can be shortened up to 12 hours. For the yeast cells, the cell cycle is much shorter and takes 1-2 hours.
Cancers are diseases characterized by the abnormal cell growth, with broken mechanisms of the cell cycle control. It is one of the key "Hallmarks of cancer" (D.Hanahan,
R.Weinberg) \cite{Hanahan2000TheHO}, \cite{Hanahan2011}, \cite{Hanahan2022HallmarksOC}. 

Classical books are devoted or have chapters on the cell cycle: 
\cite{alberts2017molecular}, \cite{weinberg2014biology}, \cite{morgan2007cell}.
The topic is still in active research. The mechanisms of most chemotherapy and target therapy types are related to deregulation of the cell cycle machinery. For example, CDK-inhibitors were the focus of multiple recent studies \cite{Asghar2017}, \cite{Fassl2022}.

\subsection{Tasks for cell cycle analysis based on scRNA-seq data \label{sectBriefOverview} }

\begin{figure}[t]
  \includegraphics[width=\textwidth]{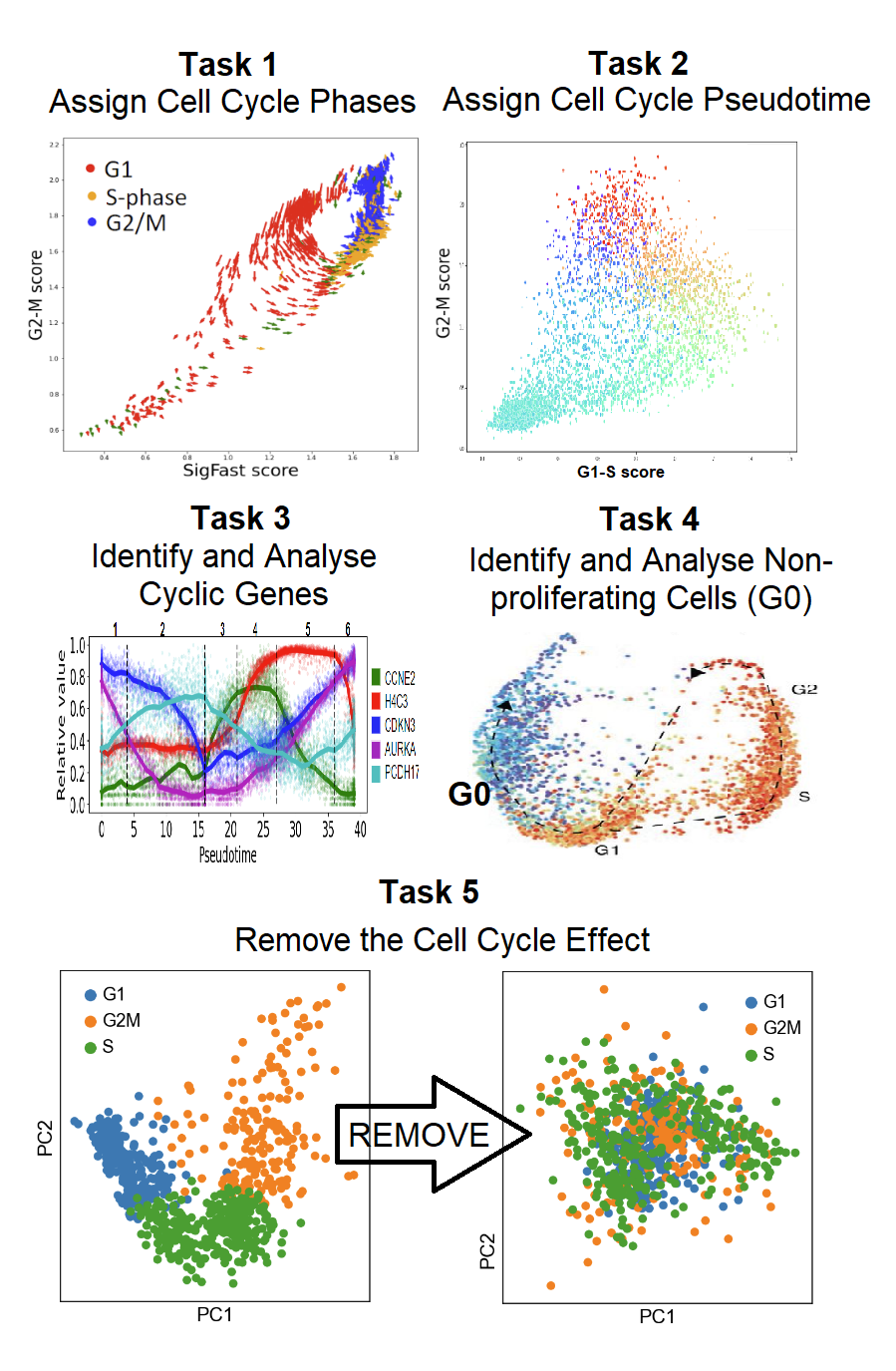}
  \caption{Five major tasks in analyzing the dynamics of the cell cycle from the single cell data.
  {\bf Task 1.} Assign the cell cycle (sub)phases. 
  {\bf Task 2.} Quantifty the cell cycle (pseudo)time. 
  {\bf Task 3.} Identify and analyse cell cycle-associated genes. 
  {\bf Task 4.} Identify and analyse non-proliferating cells (G0). (Figure from \cite{Stallaert2022}.) 
  {\bf Task 5.} Remove the cell cycle effect. (Figure from Scanpy cell cycle tutorial \cite{ScanpyCellCycle}.)
  }
  \label{fig:FiveTasks}
\end{figure}

Here we briefly introduce the computational tasks in the analysis of cell cycle dynamics from single cell data where the input is a matrix (cells x genes) and the output is one of the following (Figure~\ref{fig:FiveTasks}):

1) "Cell cycle phase labels". Each cell in the dataset should receive a label like G1,S,G2,M, G0. 

2) "Pseudotime quantification". Each cell should be attributed a real number reflecting how far the cell progressed from its birth to the cell division moment.

3) "Listing the cell cycle-related genes". For each gene one would like to quantify the statistical significance of the consistent pattern of the expression dynamics along the cell cycle trajectory and visualize this pattern.


4) "Determining G0 (quiescent) state." Distinguish  proliferating and non-proliferating (staying in quiescent state G0)  cells. In more detail, analyse the composition of the quiescent cell state and characterize the transition between proliferating and quiescent state(s).  

5) "Removing the cell cycle effect from the scRNA-seq dataset". In many studies aiming at identifying cell types, clustering cell states, cell trajectory inference, cell cycle-related heterogeneity might represent a confounding effect. The nature of this task is to remove this signal from the data.


Note that in all of these tasks we do not assume any available cell labeling, which can allow some additional types of analyses. Indeed, for example cell cycle phase labels might be provided for individual cells but this information requires a complex experimental setting and, therefore, rarely available.


The aim of this  review is to discuss in some detail these five tasks and what available packages can achieve in solving them.




\begin{table}[!ht]
    \centering
    \begin{tabular}{|l|l|l|l|l|l|l|l|}
    \hline
        Package & Lang & Year & \rotatebox[origin=c]{270}{CC Phase} & \rotatebox[origin=c]{270}{Pseudotime} & \rotatebox[origin=c]{270}{Cyclic Genes} & \rotatebox[origin=c]{270}{Remove CC} & G0 \\ \hline
        \multicolumn{8}{|c|}{Early development} \\ \hline
        Seurat/Scanpy & R/Python & 2018 & + & -- & -- & + & -- \\ \hline
        scLVM & R,Python & 2015 & -- & -- & -- & + & -- \\ \hline
        f-scLVM(Slalom) & R,Python & 2017 & -- & -- & + & + & -- \\ \hline
        Cyclone & R,Python & 2015 & + & -- & -- & -- & -- \\ \hline
        Oscope & R & 2015 & --  & + & + & -- & -- \\ \hline
        ccRemover & R & 2016 & -- & -- & -- & + & -- \\ \hline
        reCAT & R & 2017 & + & + & + & -- & + \\ \hline
        cycleX & R+Python & 2017 & + & + & -- & -- & -- \\ \hline
        \multicolumn{8}{|c|}{Recent development} \\ \hline
        Pre-Phaser & Python+C++ & 2019 & + & + & -- & -- & + \\ \hline
        Cyclum & Python+TF & 2019 & + & + & + & + & -- \\ \hline
        Revelio & R & 2019 & +  & + & + & + & -- \\ \hline
        CCAF & Python & 2020 & + & -- & -- & -- & + \\ \hline
        Peco & R & 2020 & -- & + & + & -- & -- \\ \hline
        SC1CC & R,Web & 2020 & + & + & -- & -- & + \\ \hline
        DeepCycle & Python & 2021 & + & + & + & -- & -- \\ \hline
        Tricycle & R & 2021 & + & + & + & + & -- \\ \hline
        CCPE & M+R+Python & 2021 & + & + & + & + & -- \\ \hline
    \end{tabular}
    \caption{List of computational tools for quantifying cell cycle from scRNA-Seq. Pluses and minuses indicate which tasks the tool can perform: "CC Phase" - assigning the cell cycle phase label, Pseudotime - inferring pseudotime and producing pseudotemporal plots, "Cyclic genes" - analysing/identifying genes associated to cell cycle, "Remove CC" - removing the cell cycle signal from the data, "G0" - distinguish proliferating and non-proliferating cells. The "Lang" column indicates the programming languages used for implementation. "R,Python" means that there exist two implementations, in R and Python, while "R+Python" means that part of the code is implemented in R while another part in Python.
    }
\end{table}

{\bf Computational tools assessed in this review.} 
We identified 15 software packages specifically devoted to cell cycle analysis using scRNA-seq data. 
There exist other more general scRNA-seq data analysis tools which can be applied to the cell cycle analysis. We only briefly mention scLVM and f-scLVM (Slalom) from that category together with the most popular Seurat \cite{Butler2018seuratV2} and Scanpy \cite{wolf2018scanpy} (which both have cell cycle specific tools: \cite{SeuratCellCycle}, \cite{ScanpyCellCycle} based on the same algorithm). 
There exist multiple methods for analysing periodic dependencies, not specific to cell cycle scRNA-seq data. They are also out of the scope of the present review, but we can mention the CYCLOPS package here that potentially can be used \cite{Anafi2017CYCLOPSRH,Ruben2018}.

The existing scRNA-seq datasets are quite diverse in terms of quality and the number of cells. In some high quality datasets, the cell cycle structure
can be clearly resolved by most of the tools. However, we found that none of the existing packages can uniformly well perform all the aforementioned tasks in complex/low quality scRNASeq datasets or assess the confidence of the produced results. This review aims at pointing to potential pitfalls and misinterpretations and provides some guidance in these cases.

{\bf Complex scenario, re-analysis of existing data}. 
The tasks introduced above can be solved universally only if a computational tool is aware of various possible types of cell cycle dynamics that one can observe by re-analysing hundreds of scRNA-Seq datasets. By doing such a meta-analysis, we could conclude that the transcriptomic dynamics of cell cycle can be classified in several types. There exists a well-known pattern of cyclin time dependence (Figure~\ref{fig:StandardCellCycleWithExplanations}A) which can be indeed observed in many  cell types. However, in many others, such as embrionic stem cells (ESC) and many cancer cell lines not only cyclins, but the large groups of genes
reflected in the standard signatures of G1/S and G2/M cell cycle phases can form a less known "seesaw" pattern (Figure~\ref{fig:cell_cycle_seesaw}):
when one group of genes goes up, the other one goes down and vice versa. This means that immediately after the mitosis, G1/S signature genes begin to increase their expression. In this mode the restriction point (R-point), which is normally characterized by some minimal expression level of most cell cycle genes simultaneously, is not observed as a transcriptomic state. This is in line with the well-known observation of the G1-phase shortening for ESC 
(e.g. \cite{becker2006self}, \cite{neganova2008review}, \cite{kapinas2013abbreviated}, \cite{coronado2013short}, \cite{Kolodziejczyk2015}, \cite{DeepCycle}, etc.), reflected at the level of single cell transcriptomes. Thus, the anticorrelation of G1/S and G2/M signature score is an indication  of the "fast" (ESC-like) proliferation. 

We observe such pattern for many cancer cell lines. 
Biologically this can be related to the deregulation of the TP53-p21(CDNKN1A)-Cyclin/CDK-Rb-EF pathway (e.g. \cite{Engeland2022}). We also observe that many computational tools aiming at the analysis of cell cycle dynamics from single cell data (including the most popular one such as Seurat/Scanpy) are not able to interpret such pattern correctly. We explain the nature of this difficulty in detail and suggest ways of dealing with it (see subsection \ref{sectFastCC} ).

Even more complex phenomena may sometimes appear: within even seemingly homogeneous cell populations, e.g. cell lines, some cells may proliferate according to the "fast" pattern, some according to the "standard" pattern. We observed such a pattern in several well-known cell lines profiled at single cell level such as U2OS or MCF10-2A. Such phenomenon has not been described yet in scRNA-seq related literature before and none of the packages can correctly deal with it. The situation is especially difficult since the "fast" and the "standard" cell cycle trajectories are merged as "Siamese twins", having the common "S-G2-M" segment with only G1 part being different (see Figure~\ref{fig:doubleCellCycle}C). We demonstrate possible approaches to resolve this issue (see subsection \ref{sectDoubleCC}). 
The existence of the bifurcation in the cell cycle dynamics is in principle described in the literature \cite{Spencer2013}, \cite{Matson2017}, \cite{Min2019}. Moreover, it can be related to the existence of the minor cancer cell subpopulations resistant to the treatment by the CDK-inhibitors \cite{Asghar2017}. These studies were done using proteomics and image analysis-based approaches targeted on specific proteins such as CDK2 and Cyclin E.
Observing such a bifurcation pattern at the level of scRNA-Seq provides new opportunities for the mechanistic understanding of its nature, using genome-wide transcriptome analysis, which principles are explained in the section \ref{basicIdeas}. 

The review grow out of developing our own methodology to analyse the  cell cycle using single cell RNA sequencing data,
a part of it can be found in \cite{aynaud2020Ewing}, \cite{zinovyev2021modeling}, which is based on ElPiGraph package \cite{albergante2020robust}.
We have analysed several hundreds datasets in particular from  collections like ARCHS4 \cite{lachmann2018archs4},
and atlas scale "Tabula Muris" \cite{tabula2018single} , "Tabula Sapiens" \cite{tabula2022Sapiens}, "ENCODE" \cite{He2020}, etc.
Analysis shows that for the standard cases of the cell cycle many approaches work quite fine,
but as described above there are more subtle situations where analysis is challenging.
We put 50+ datasets and notebooks based on our own methodology on the Kaggle cloud service,
which is distinguished in that it allows one to execute code and  store data in one place. 
It can be found at:  \href{https://www.kaggle.com/search?q=scRNA-seq+in\%3Adatasets}{https://www.kaggle.com/search?q=scRNA-seq+in\%3Adatasets}.

\section{Studying the cell cycle transcriptomic dynamics at genome-wide level \label{sect_scrnaseq_vs_other}}

Let us provide a brief reminder on some biological cell cycle studies related to the current review. 
The first subsection concerns more classical studies, and the second single cell RNA sequencing ones.

\subsection{Synchronized and asynchronized cell cycle studies  \label{sectBulkTranscriptomicStudies} }

Cell cycle has been widely studied long before the appearance of single cell technologies. Early studies that shaped the understanding of the main principles of the cell cycle (in particular Nobel prize wining) were based on studying expression dynamics of a restricted number of proteins. 


Genome-wide transcriptomics studies of yeast cell cycle began to appear around 1998\cite{spellman1998comprehensive} followed by the seminal studies of cell cycle in HeLa human  cell line \cite{Whitfield2002}. 
In late 2000s, several systematic genome wide transcriptomic studies were made using other human cell types, human foreskin fibroblasts \cite{BarJoseph2008}, an immortalized human keratinocyte cell line (HaCat) \cite{PenaDiaz2013}, and the osteosarcoma-derived cell line (U2OS) \cite{Grant2013}. These studies reported 874, 480, 1249, 1871 cell cycle related genes. The following meta-analysis revealed certain difficulties to draw a fully consistent picture of the cell cycle \cite{Grant2013}, \cite{Giotti2017}, \cite{Giotti2019}.
Firstly, there was a rather small intersection in the gene lists: "despite the pairwise overlaps being in the 40 percent range, there are 142 genes that are cell cycle regulated in all three cell types" (and adding the 4th dataset one gets only 96 genes) \cite{Grant2013}. Secondly, the assignment of gene activities into cell cycle phases (based on the peaks of expression) was also not consistent: "only 18 genes were annotated as G2/M and 16 genes as G1/S consistently across all four studies, while for the other phases (S, G2, M/G1), not even a single gene was identified by all studies" \cite{Giotti2017}. 

Studying the cell cycle based on bulk transcriptome relies on the cell synchronization, i.e. cell populations must be blocked in a certain phase of the cell cycle (e.g. by double thymidine block,  by thymidine-nocodazole block, etc), then simultaneously released from the block, and some fractions of cells are longitudinally sequenced. Thus one can obtain the temporal plots of the genes' expression dynamics. Examples of such plots can be found at the Cyclebase website \cite{Cyclebase3}
for HeLa and several other cell lines. These experiments do not need single cell sequencing because we can assume that all the cells are in the same position of the cell cycle progression: however, this assumption is known to be not fully realistic. Cells progress the cell cycle not exactly with the same speed and some do not progress at all, thus synchronization is not exact. To overcome this issue several single cell (even though not scRNA-seq) experiments have been performed, see them reviewed in \cite{Matson2017}. 

Importantly, these early works revealed certain deviations from the classical textbook picture of the cell cycle. In particular, it has been discovered that progression through the R-restriction point is not required for all cells. In some situations there were cell subpopulations (within a seemingly homogeneous cell line) which progressed through the cell without R-point \cite{Spencer2013}, \cite{Matson2017}, \cite{Min2019}. They were committed to the next round of cell cycle during the preceding round, and they could progress through the cell cycle in a mitogen independent way.  


Interestingly, in early times it was thought that all the CDKs are equally important for cell cycle progression in all cells. However, later it was understood that certain CDK are only essential for proliferation of specialized cells. Thus, it was hypothesized that "selective CDK inhibition may provide therapeutic benefit against certain human neoplasias" \cite{Malumbres2009ChangingParadigm},  \cite{Gordon2018Cellcyclecheckpointcontrol},  \cite{Matthews2022Cellcyclecontrolincancer}.  

Another important research direction is studying the effect of knocking out cell cycle-related genes and estimating their impact on the cell cycle phenotype
\cite{Mukherji2006}, \cite{Sokolova2017}, 
\cite{VinerBreuer2019}. For example, using this approach one can compare cancer and non-cancer cell lines and emphasize the difference in their cell cycle regulation. 

Of course, there are great numbers of other important cell cycle studies that we cannot mention all here. 


\subsection{Studying the cell cycle at single cell transcriptome level \label{sect_cellcycle_scrnaseq}    }


ScRNA-seq method provided a huge impact on many fields of biological research. In particular, the structure of cell populations can now be studied much more directly than ever.  New small subpopulations have been discovered, differentiation and other dynamical processes like epithelial-to-mesenchymal transition (EMT) can be understood much better now with the use of single cell approaches. Computational methods play a pivotal role in these studies \cite{Lahnemann2020},  \cite{Kiselev2019},
\cite{Bergen2021}, \cite{Kharchenko2021}. 

Cell cycle is a major focus in many scRNA-seq studies. It is important in the context of cancer biology since proliferation is one of the key cancer hallmarks \cite{Tirosh2016}, \cite{Tirosh2016b}, \cite{Kinker2020}. A connection between major oncogene drivers and the cell cycle progression has been studied at single cell level (for example, see \cite{aynaud2020Ewing}). A typical problem in cancer treatment is the appearance of drug resistance which is sometimes attributed to the existence of small subpopulations of tumor cells which are able to survive the treatment. Therefore, scRNA-seq methods appear natural for searching,  characterizing such subpopulations and studying their adaptation strategies \cite{Zinovyev2021PhysOfLife}. One of the ways to attack the resistance problem is to use combinations of drugs.
Recent single cell studies like \cite{Aissa2021} suggest ways for how to search for such combinations involving the analysis of cell cycle. 
Another example:  minor cell subpopulations in prostate cancer that are androgen independent are the possible reasons for the resistance to  androgen-deprivation therapy. Using single cell-based analysis of the cell cycle it was suggested that they can  characterized by the  enhanced expression of 10 cell cycle-related genes: CCNB2, DLGAP5, CENPF, CENPE, MKI67, PTTG1, CDC20, PLK1, HMMR, and CCNB1 \cite{Horning2018}.


Single cell approaches give a hope to decipher the connection between the genotype and complex phenotypes such as cell cycle. Recent breakthrough technologies greatly upscale that opportunity:  thus, the Perturb-seq protocol merges CRISPR with scRNA-seq and provides the data on systematic gene perturbations on unprecedented scale \cite{REPLOGLE2022}, \cite{Ursu2022}.  For example, it was reported that "the proportions of cells in each cell cycle phase was highly accurate (area under the precision-recall curve of 1), suggesting that  our TP53 variant phenotypes can be predicted from scRNA-seq data" \cite{Ursu2022}.  In what follows we will also discuss some other relation between TP53 and cell cycle dynamics type. 

{\bf Pluripotent stem cells, relation of the cell cycle and differentiation.} For this review it is important to keep in mind that embryonic stem cells (ESC) are known to have an unusual cell cycle pattern with shortened G1. There are numerous reports on the modified properties of cell cycle in ESC and iPSC cells. It appears important to understand whether the unusual cell cycle machinery is related to  sustaining pluripotency (probably yes, but to what extent?) and, if yes, what is the relation between pluripotency and cell cycle machinery. 

In particular, it has been argued that differentiation typically happens in G1, so to sustain the pluripotency, it is desirable to keep G1 as short as possible that can explain short G1 for ESC.  ScRNA-seq studies begin to play a role in such questions \cite{Natarajan2017}, \cite{Liu2019}. 

In one of the earliest scRNA-seq study,  mESC cells were assessed in different conditions, reporting
differential heterogeneity in the cell cycle genes between conditions and a link with the cell cycle speed \cite{Kolodziejczyk2015}.
In another study, use of scRNA-seq methods revealed that the fate of differentiation (either to extraembryonic endoderm cells (XENs) or to epiblast stem cells) depends on the cell cycle phase in which the differentiation stimuli (retinoic acid-based protocol) was applied \cite{Levi2020}.

{\bf G0, quiescence, senescence, cell cycle exit and reentry.}
Understanding the cell cycle machinery is intimately related to understanding how the cells exit and (re)enter the cell cycle from/to G0 (senescence, quiescence) state. Understanding the cellular dynamics within quiescence and senescence state is also of great importance.  That has been a topic of many studies for dozens of years \cite{Cheung2013} but many questions remain to be understood. 

Let us mention a few scRNA-seq related studies that are related to the present review.

In one of the earliest and excellent scRNA-seq papers, where many ideas of this review had appeared, the single cell dynamics of hematopoetic cell aging was studied in mouse \cite{Kowalczyk2015}. The conclusion was tightly related to the cell cycle analysis: it was found that "cell cycle dominates the variability within each population and that there is a lower frequency of cells in the G1 phase among old compared with young long-term HSCs, suggesting that they traverse through G1 faster". It is interesting to note that age differences were not observed among some non-cycling populations like multipotent progenitors. 

CCAF is a Python package with  "a cell cycle classifier that identifies traditional cell cycle phases and a putative quiescent-like state in neuroepithelial-derived cell types during mammalian neurogenesis and in gliomas" \cite{CCAF}. The authors mention that knockouts of "Hippo/Yap and p53 pathways diminished Neural G0 in vitro, resulting in faster G1 transit". 
The paper proposes many other interesting insights to molecular mechanisms of senescence in the context of neural cells, and it might be that at least part of these findings can be generalized to other cell types. The study has a potential clinical importance because quiescent glioma stem cells might be a reason for the treatment resistance. 

Recently an innovative approach for estimating cell cycle phases and pseudotime was suggested \cite{DeepCycle} - based on studying the balance between spliced and unspliced transcript expression (idea used in RNA-velocity \cite{Bergen2021}).
The structure of a human fibroblast subpolulation at G0 (or nearby) was documented.  The paper analyses several published gene markers for quantifying quiscence and shows that their results are consistent with the expectations. 

Let us also mention a recent single cell proteomics
study \cite{Stallaert2022}, \cite{stallaert2022molecular}
where the single cell measurement of expression of 48 proteins allowed the authors to characterize the structure of the senescence state, score the "degree of senescence", and propose non-classical ways of exit from the cell cycle (from G2 to G0). An interesting question is whether these results can be reproduced using the scRNA-seq methods. 

To conclude, scRNA-seq is a powerful technology which enabled remarkable progress in many directions. Even though today the main focus of the scRNA-seq studies is not on the cell cycle, but on subpopulation studies, differentiation etc, already obtained results suggest that the scRNA-seq approach promises to substantially improve our understanding of the cell cycle regulation. This might be important for many clinically relevant questions such as those related to CDK-inhibitors \cite{Suski2021}, \cite{Fassl2022} in cancer research and other domains.  Let us also note that single cell studies (not RNA-seq, but mainly based on image-based protein expression measurements) provided important insights on the cell cycle starting from  almost ten years ago \cite{Spencer2013}, \cite{Matson2017},\cite{Asghar2017}.

\subsection{Advantages and disadvantages of single cell transcriptomic methods \label{sectAdvDisadvSCRNAseq} }

scRNA-seq methods have certain limitations: they can only measure the transcriptome. Expression of proteins and their post-translational modifications are important for studying the cell cycle. Compared to the bulk transcriptome experiments with cell synchronization (as those mentioned above) scRNA-seq typically produces more noisy data and, thus, less precise. E.g. most recent synchronization methods report up to 20 minutes time resolution of gene expressions profiles \cite{Chervova2022} that is quite beyond the current scRNA-seq resolution. Single cell measurements are also substantially more expensive than bulk methods. However, scRNA-seq methods are already produced and keep producing large amounts of publicly available data, for various cell types, cell lines and perturbations. For example, the transcriptomic profiles of 700 000+ cells from the three cell lines exposed to more than 180 drugs were produced in a single screening \cite{Srivatsan2020}.
Bulk experiments explored in great detail 4 cell lines in 10 years (as discussed above) while the scRNA-seq can produce results on hundreds of cell lines in one study (such as the collection of CCLE with ~200 cancer cell lines from 22 cancer types) \cite{Kinker2020}. 
Not all of the available single cell datasets contain the profiles of proliferating cells, but many of them do. Let us emphasize that any scRNA-seq dataset containing proliferating cells can be used for the cell cycle analysis, i.e. there is no need of any special experimental design such as cell synchronization, 
no use of specific protein markers. The analysis is purely computational: thus it can be readily done for already published datasets. 




%

\section{Basic methods of the cell cycle analysis using  scRNA-seq data, pitfalls and guidelines \label{basicIdeas} }

Before going into the details in the next sections, here we 
will try to explain in simple words the main ideas and subtleties of the cell cycle analysis based on scRNA-seq data.
We also decided to include some our original findings
(which we will elaborate in subsequent publications), because
they seems to clarify certain points as well as more explicitly
highlight subtle situations of the analysis. 


\subsection{G1/S-G2/M-phase plot - the first, simple and a must thing to do \label{sectG1SG2MplotFirst} }

The first and simple thing in the analysis of cell cycle using scRNA-Seq data is to create the G1/S-G2/M-phase data visualization. It is interpretable and allows the visual control of the analysis. We assume that log-based transformation and normalization have been already performed for scRNA-seq dataset.

{\bf Plot construction.} The plot is based on visualizing the scores of G1/S and G2/M gene signatures, such as it was first proposed in \cite{Tirosh2016}.  The score can be the mean of the expression levels for the genes in a signature, or other suitable score. Therefore, each cell is characterized by two numbers and the phase plot represents a simple scatterplot of all cells in such two dimensional plane. 

\begin{figure}[t]
  \includegraphics[width=\textwidth]{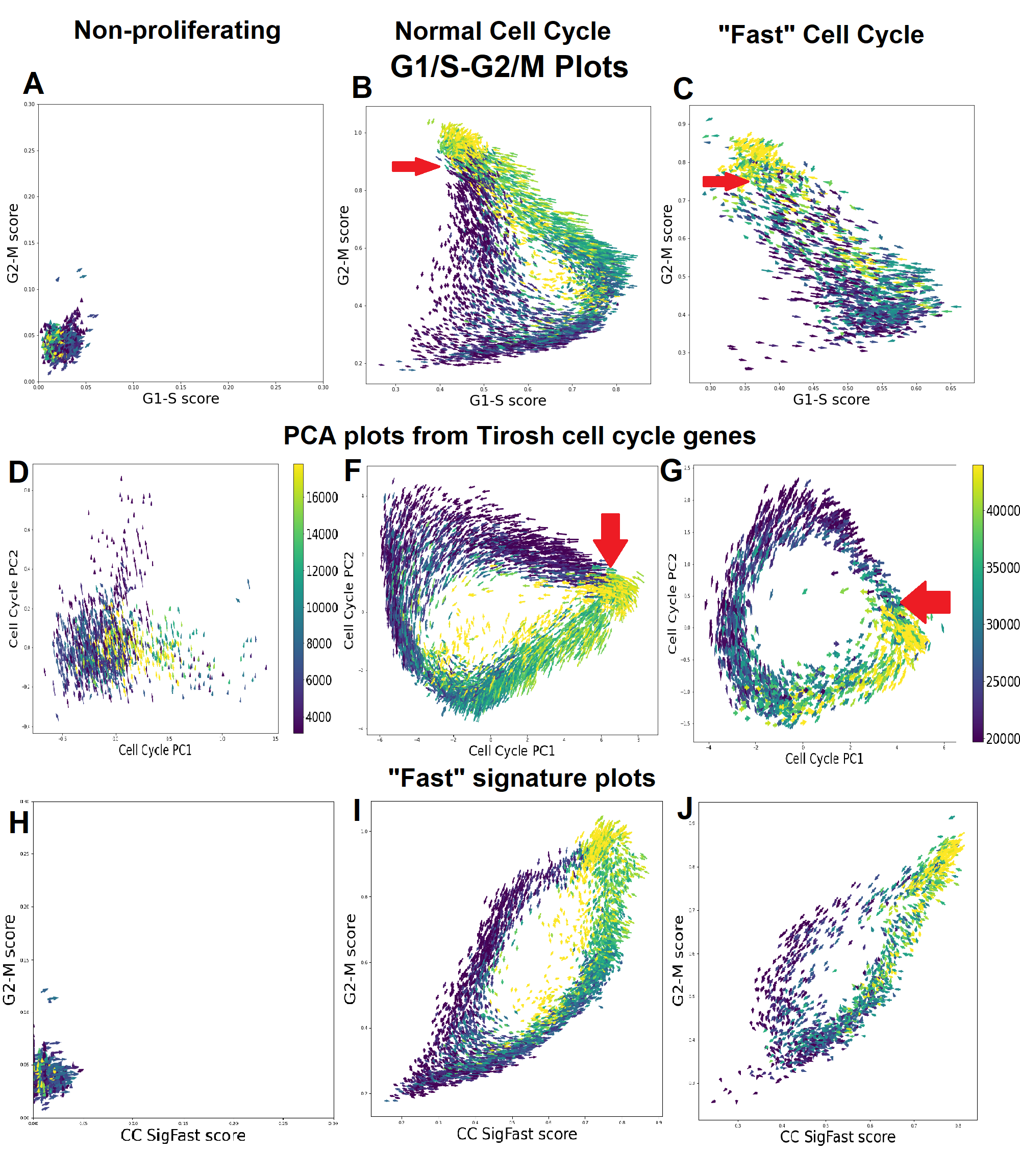} 
  \caption{Three patterns of the cell cycle tjajectory seen at the level of transcriptome. Three ways of visualization.
  Panels in the first column correspond to non-proliferating cells (PMBC dataset); the second column corresponds to proliferating cells with the "standard" cell cycle pattern (CHLA9 cell line); the third column corresponds to the "fast" cell cycle pattern (TC71 cell line),
  which is  typical for ESC, iPSC, cancer cells with TP53 mutation.
  Panels in the first row correspond to G1/S-G2/M visualisation;
  the second row to the PCA visualization is restricted to cell cycle genes from the Tirosh's set; the third row corresponds to the special gene signature designed to visualize the fast cell cycle pattern (see section \ref{sectFastCC}). The red arrows mark the approximate position of the mitosis end. Color maps the total number of read counts per cell: a trend with maximum at mitosis is clear. Each cell is visualized with an arrow representing the RNA-velocity vector, its direction and amplitude.
  }
  \label{fig:review3typesOfCellCycle}
\end{figure}

{\bf Interpretation step 1, classifying the type of cell cycle pattern (standard vs non-proliferating vs ESC-like ("fast"))}.
As our analysis of hundreds datasets shows, the obtained plot may typically fit in one of the three patterns (Figure~\ref{fig:review3typesOfCellCycle} upper row): 

0) the one of a population of non-proliferating cells - Figure~\ref{fig:review3typesOfCellCycle}A (of note, most of the normal cells of an adult ogranism are not proliferating, this is a widespread pattern).

1) the one corresponding to the "standard" cell cycle typical for normal and some of the cancer cells - Figure~\ref{fig:review3typesOfCellCycle}B. Typically, it represents a triangle-like shape which we will discuss in detail below.

2) the "fast" pattern - Figure~\ref{fig:review3typesOfCellCycle}C - characteristic for fast proliferating cells such as ESC, iPSC, certain cancer cell lines, probably to T-cell at the stage of the clonal expansions (compare with \cite{Singh2022}). One can see that the standard G1/S-G2/M plot is difficult to interpret in this case. We will discuss how to deal with it below, and for the moment it is important to note that the G1/S-G2/M plot easily allows one to  distinguish such type of the cell cycle by mere visual inspection. 


{\bf History of G1/S-G2/M plots.} G1/S-G2/M plots were first introduced (to the best of our knowledge) around 2015 by Tirosh, Kowalczyk, Regev et.al.: see Figure 2F in \cite{Kowalczyk2015}, Figure 2A in \cite{Tirosh2016}, Figure 3A in \cite{Tirosh2016b}. Firstly they were suggested on the basis of Whitfield's cell cycle phase gene signatures \cite{Whitfield2002}. Later on this approach became standard in Seurat/Scanpy packages. However, at that time the datasets were quite small and noisy as one can see on the cited figures. The triangle-like pattern and its interpretation (see the next section) seems to be not clearly stated in the literature. It became possible because the quality and the size of the single cell datasets improved over the recent years. Another possible reason which probably prevented some of the authors from using this type of visualization is the existence of the "fast" cell cycle pattern, which might seem uninterpretable in G1/S-G2/M plot. 

{\bf G1/S-G2/M plots in the standard packages (Seurat/Scanpy)}. The simplest way to compute G1/S and G2/M scores for a single cell is to use the mean values of normalized and logged gene expression in the signature. Some of the standard packages use an improvement of this simplest approach by scaling and applying reference-based transformation of the gene expression levels. From our experience, such transformation does not change the  interpretation of the cell cycle trajectory based on the analysis of its segments. At the same time, the appearance of negative score values and the non-unique choice of the reference gene sets for the transformation can make the interpretation of the standard G1/S-G2/M plot more difficult.

\subsection{Interpretation of the  "standard" ("triangle") cell cycle pattern. \label{InterpretationOfTriangle} } 
\begin{figure}[t]
  \includegraphics[width=\textwidth]{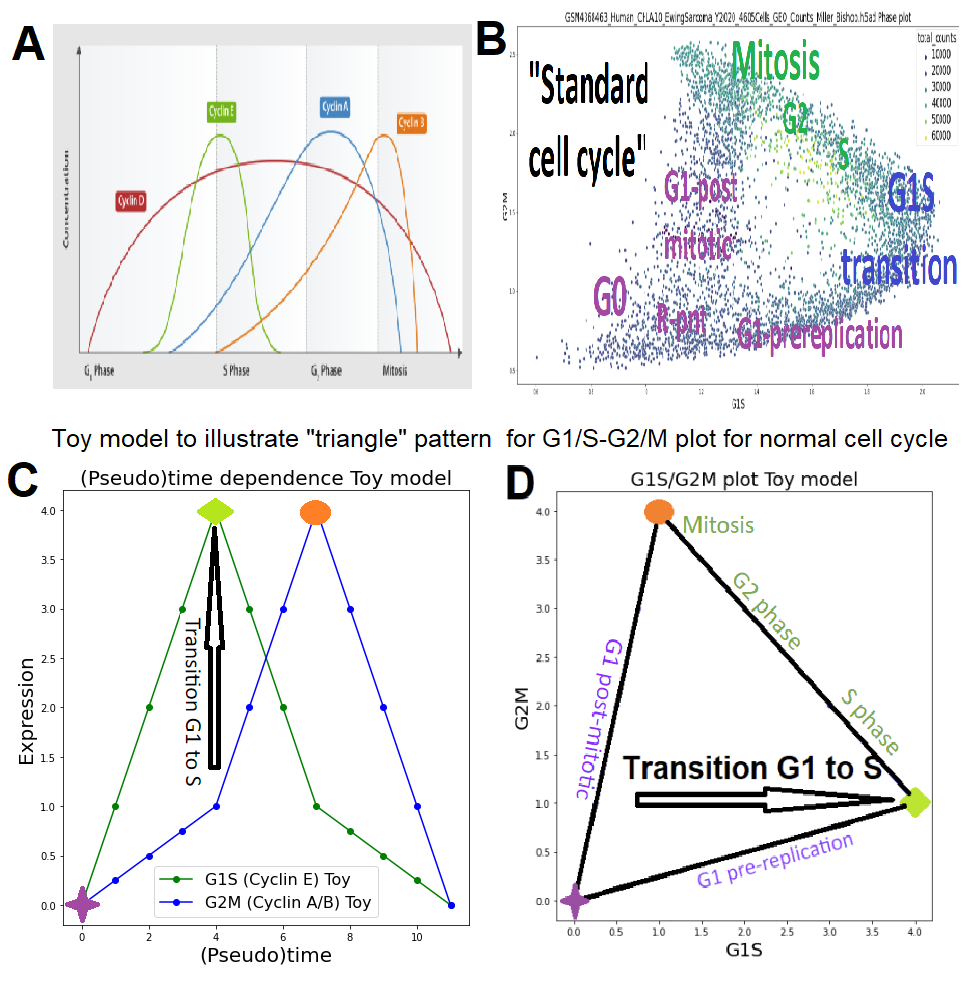} 
  \caption{The "standard" cell cycle trajectory, its interpretation and the toy model.
  {\bf A.} Well-known plot of cyclins activities.
  {\bf B.} The G1/S-G2/M plot typical for normal proliferating cells - "triangle" pattern.
  {\bf C.} Toy model for cyclin activities. 
  {\bf D.} G1/S-G2/M plot corresponding to the toy model.  
  }
  \label{fig:StandardCellCycleWithExplanations}
\end{figure}

The G1/S-G2/M plot for the normal proliferating cells typically  looks like a "triangle" (see Figure \ref{fig:StandardCellCycleWithExplanations}B). It is just an approximation to the observed shape, but convenient to start with.

{\bf Left vertex: "G-zero near zero".} The left bottom corner (near zero of coordinates) - corresponds to cells which are either non-proliferating or in such part of G1 which is transcriptionally similar to non-proliferating cells ("G0"-state). 

{\bf Top vertex: "Mitosis is on top".} The top corner of the triangle corresponds to mitosis. That is the maximum of the G2M group of genes in particular cyclins A,B whose maximum is approximately near mitosis. One should keep in mind that mitosis takes only small part of duration  of the cell cycle, 
so in typical scRNA-seq dataset only a small fraction of cells is in mitosis.  And so it is difficult to place very exact bound between G2,G1  phases and mitosis with scRNA-seq data,
but its rough position is quite clear: it is located near the top corner of the triangle. 

{\bf Left edge: "G1-postmitotic". } The left segment of the triangle (from top to bottom-left) corresponds 
to the post-mitotic part of G1-phase. That part starts after the mitosis and ends at the R-(restriction point), where the cell decides to go to G0 or to the next round of the cell cycle.

{\bf Bottom edge: "G1-prereplication". } The bottom segment of the triangle corresponds to the second part of G1 phase
(pre-replication) where cells committed to the next round of the cell cycle are starting their preparation for the S-phase.

{\bf Right vertex: "G1/S transition". } The right corner of the triangle corresponds to the transition from G1 phase to S-phase. That is the maximum of the G1/S group of genes, in particular, cyclin E (CCNE1, CCNE2).

{\bf Right edge: "S-G2-M". } The right segment of the triangle corresponds to the three phases in once: S-phase,G2-phase and a part of M-phase.
As mentioned above mitosis is quite short, so it is not surprising that it is hardly distinguishable from G2.
Tt is more surprising that the bound between S-phase and  G2 phase  is not highlighted by something like a corner
of the shape. We will address that question later (one can probably choose other gene signatures rather
than G1/S, G2/M and it might be possible to observe the change in the transcriptome dynamics).

{\bf Rationale behind the interpretation of "triangle" for the "standard" cell cycle and subtle points.} The rationale behind the "triangle" interpretation is simple and based on the textbook plots of cyclins activities (Figure \ref{fig:StandardCellCycleWithExplanations}A). We can consider a simplified toy model for the G1/S gene set just to consist from only cyclin E, and G2/M gene set can be representd by cyclin A or B or their average (Figure \ref{fig:StandardCellCycleWithExplanations}C). Then
the classsical textbook plots for their activities would result in that the shape of the G1/S-G2/M phase plot would be 
close to a triangle-like shape (Figure \ref{fig:StandardCellCycleWithExplanations}D), with the vertices and edges of the triangle corresponding to  what has been described above. 
The step from the toy model to real plot is also conceptually simple: the point is that there are many genes which are correlated to cyclin E, and they are gathered in the G1/S signature. There is another gene set G2/M collecting the genes correlated to the cyclins A and B.  In the other words, genes from the G1/S peak near the border of G1 and S phase, and genes from the G2/M peak in  mitosis. Therefore, making a 2D plot based on the means of the two gene sets, one obtains a triangle. Taking the means of relatively large gene sets increases the signal to noise ratio and makes the triangle-like shape more prominent.

More formalized model of the triangular-like shape of the standard transcriptomic cell cycle trajectory in the G1/S-G2/M phase space was suggested recently in \cite{zinovyev2021modeling}. The modeling is based on the general dynamical theory of self-replicating systems as allometric growth with switches and it allows one to infer some ratios between physical time durations of the cell cycle phases.


The biological reason for the existence of correlated sets of genes is also clear: G1/S gene set is co-regulated by the powerful transcription factor family of E2Fs, while the G2/M gene sets might be also co-regulated, in particular, by the FOXM1 transcription factor. Potent transcription factors can activate multiple genes simultaneously, that is the reason why there are large groups of genes which behave in similar fashion.  Let us also note that similar "waves" of transcription have been reported for yeasts, see the beautiful Figure 4  from \cite{Rowicka2007}. Therefore, it is natural to think that the existence of two co-regulated gene sets G1/S and G2/M is fundamental to all eukaryotes. 

In most of our analyses, we used a selection of G1/S and G2/M markers made by Tirosh, and it produced good results. However, one should keep in mind that it is not that much important 
to take precisely these genes: if some of them are absent or low expressed, the analysis is reproducible with some parts of these sets. One can add much more genes to these groups (as, for example, in \cite{Giotti2017}, \cite{Giotti2019}): however, in our experience, adding more genes is possible, but would give higher correlation between
obtained signatures, which can impair the visualization. 
Therefore, the Tirosh gene sets are quite robust and a reasonable choice for visualization of the standard cell cycle. 

Of note, identification of gene sets for defining a cell type-specific and cell cycle-related gene signatures can be done in a (quasi-)unsupervised way. For example, it was shown that the application of Independent Component Analysis (ICA) to scRNASeq data (using the full set of genes) usually produces at least two independent components that can be matched to the known G1/S and G2/M signatures (\cite{aynaud2020Ewing,Sompairac2019} but some cell type specific genes will be highlighted in these components as well. Interestingly, matrix factorization applied to scRNA-Seq data frequently identifies more than two latent factors associated with cell cycle through gene set enrichment analysis \cite{zinovyev2021modeling, Gavish2021biorxiv}.

Yet another remark: it is known that G1/S, G2/M gene sets can be subclustered to get finer clusters of genes. We will discuss this aspect below (see again the Figure 4 in  \cite{Rowicka2007} for yeasts).

\subsection{"Fast" (ESC-like) cell cycle trajectory pattern seen through the single cell transcriptomic measurements. \label{sectFastCC} }
Let us now explain the other pattern seen on the G1/S-G2/M plot (Figure \ref{fig:review3typesOfCellCycle}C), characteristic for the "fast" dynamics.

It is well-known that embryonic stem cells proliferate faster than typical normal cells and have somewhat different pattern of the cell cycle, which is typically described in the literature as "shortened G1 phase" (e.g. \cite{becker2006self}, \cite{kapinas2013abbreviated}, \cite{coronado2013short}, \cite{Kolodziejczyk2015}, \cite{DeepCycle}, etc.). 


\begin{figure}[t]
  \includegraphics[width=\textwidth]{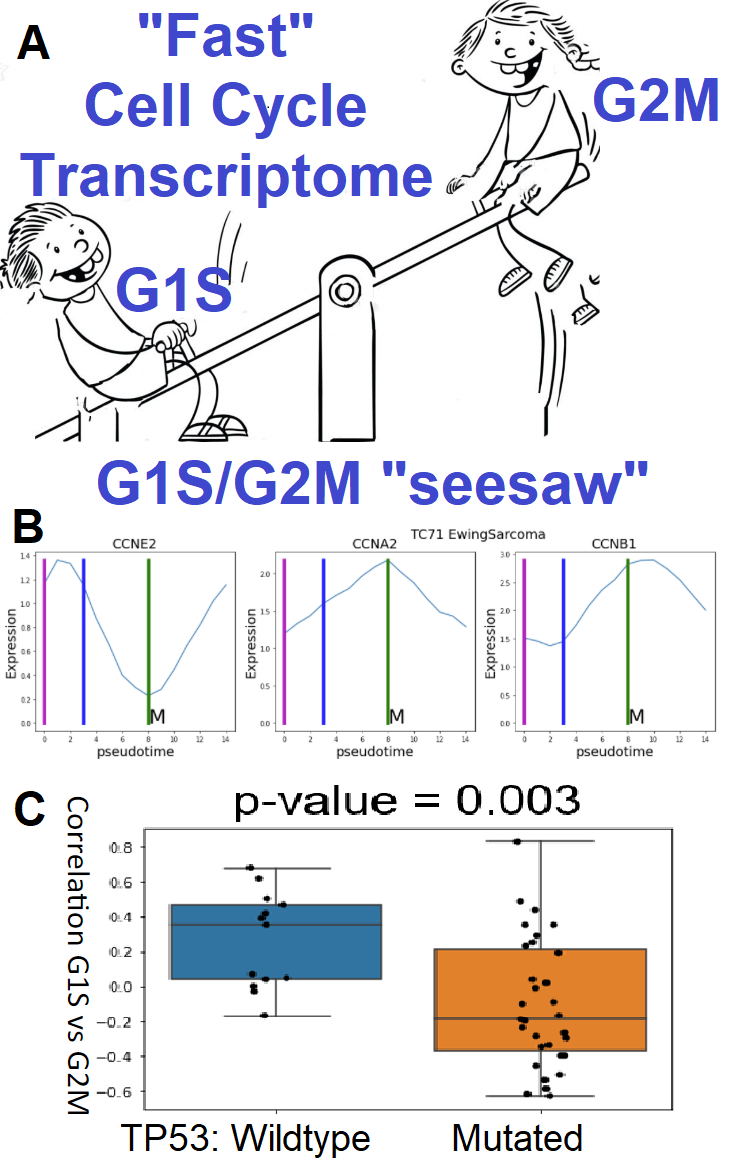} 
  \caption{The "fast" cell cycle: seesaw pattern for G1/S vs G2/M genes, relation to TP53.
  {\bf A.} Pictorial illustration for the fast cell cycle "seasaw" pattern. Expression of G1/S genes goes up, when G2/M goes down and vice versa. 
  {\bf B.} Pseudotime profile of the key cyclins for the fast cell cycle. 
  {\bf C.} The correlation of G1/S vs G2/M scores shows significant dependence on the fact of TP53 mutation for cell lines in Pan-cancer cell line heterogeneity Study \cite{Kinker2020}. 
  }
  \label{fig:cell_cycle_seesaw}
\end{figure}

The "fast" cell cycle trajectory seen at the level of single cell transcriptomes can be characterized through the following angles (see Figure~\ref{fig:cell_cycle_seesaw}):

{\bf G1/S-G2/M (pseudo)time "seesaw".} At any point of the fast cell cycle - either the G1S group of genes is growing or the G2M group of genes is growing, like sitting on the opposite sides of the children's seesaw.  There is no period where both groups are on the low level or even both are decreasing in contrast to the standard cell cycle (where both groups are decreasing after mitosis), and there is no R-point (where both groups are on minimal level). 
See Figure \ref{fig:cell_cycle_seesaw}.

{\bf G1/S-G2/M anticorrelation.} A simple computational test for the "fast" cell cycle can be suggested which consists in that the correlation between G1/S and G2/M signatures is negative or close to zero. That is a mathematical formulation of what was said above: G1/S goes down when G2/M grows and vice versa. 

{\bf G1/S-G2/M visualization} The "fast" cell cycle in G1/S-G2/M plot looks like a 2D data point consisting of a single segment from left-up (mitosis) to bottom-right (G1/S border) - Figure \ref{fig:review3typesOfCellCycle}C. 

{\bf Proliferation with no R-point.}
In the standard pattern of cell cycle trajectory, there exists a resting point, which corresponds to the R-point, where the expression of most of the cell cycle genes reach their minima. In the "fast" cell cycle trajectory pattern the G1/S group of genes begins to increase their expression immediately after the mitosis (while the G2/M group of genes goes down).


Let us emphasize that what is described above is a molecular characterization
of the cell cycle pattern, which can be not always connected to the actual "fastness" in the physical time and "ESC-likeness" of a cell type. The available data and common sense indicates that the pseudotemporal and actual physical duration should be correlated, but it is not crucially important for the analysis of the trajectory. 
Understanding the molecular mechanisms of the "fast" cell cycle, and its switch to "standard" and the relation with cancer seems to be an interesting direction of future research. 

{\bf Visualization of the "fast" cycle excluded region.}

One of the striking features of the standard cell cycle trajectory visualization is the presence of an excluded region or a region with low point density or a "hole", in the G1/S-G2/M phase plot. The existence of this region points to the presence of a cyclic process. However, for the "fast" cycle, due to the anti-correlation pattern, the excluded region is frequently not present. 

Through re-analysis of multiple "fast" cell cycle trajectories, we suggest to replace the G1/S signature with another one (that we named $CC\_SigFast$), consisting of the following genes: CDK1, UBE2C, TOP2A, TMPO, HJURP, RRM1, RAD51AP1, RRM2, CDC45, BLM, BRIP1, E2F8, H2AC20. Visualization of the single cell RNASeq data on the phase plot "CC\_SigFast vs G2/M" reveals the excluded region which serves as an indication of a cyclic process (see Figure~\ref{fig:review3typesOfCellCycle}J).



An alternative strategy to reveal the existence of excluded region in the gene expression space is to consider a set of cell cycle genes (e.g., from both Tirosh's G1/S and G2/M groups) and  make use of the standard PCA for visualization (see Figure~\ref{fig:review3typesOfCellCycle}J).
In some difficult cases (e.g. mixed cell cycle types as described below: presence of normal and fast cell cycle), one may look at PCA not in the  first two main components, but also check other pairs of higher order components.  

The disadvantage of use of PCA for visualization in comparison to the use of pre-defined gene sets is that interpretability is not straightforward. In particular, it is less easy to label different parts of the cell cycle trajectory as mitosis, G0, etc. The comparison between different cell types of conditions is also more difficult.

\subsection{Mixture of the normal and "fast" cell cycle within one cell type \label{sectDoubleCC} }


It appears to be that some datasets demonstrate unexpected phenomena -  within seemingly homogeneous population of cells, even from one cell line, some cells may proliferate according to normal pattern, while others 
according to the "fast" pattern.  
Similar phenomenon has been described in the cell cycle community: “live imaging of the fluorescent reporter in single cells revealed the two subpopulations of cells with different Cdk2 activity levels and subsequently different G1 phases”  \cite{Matson2017}, and especially in the works by S. Spencer 
\cite{Spencer2013} , \cite{Min2019}, where the role of overactivated CDK2 was described. Later \cite{Asghar2017} reported that some breast cancers cells lines have certain actively proliferating subpopulations, which  are the reasons for resistance to CDK4,6 inhibitors like Palbociclib, and characterized them by highly active CDK2 and cyclin E. 

Let us describe example with mixture of the normal and "fast" cell cycles. 

\begin{figure}[t]
  \includegraphics[width=\textwidth]{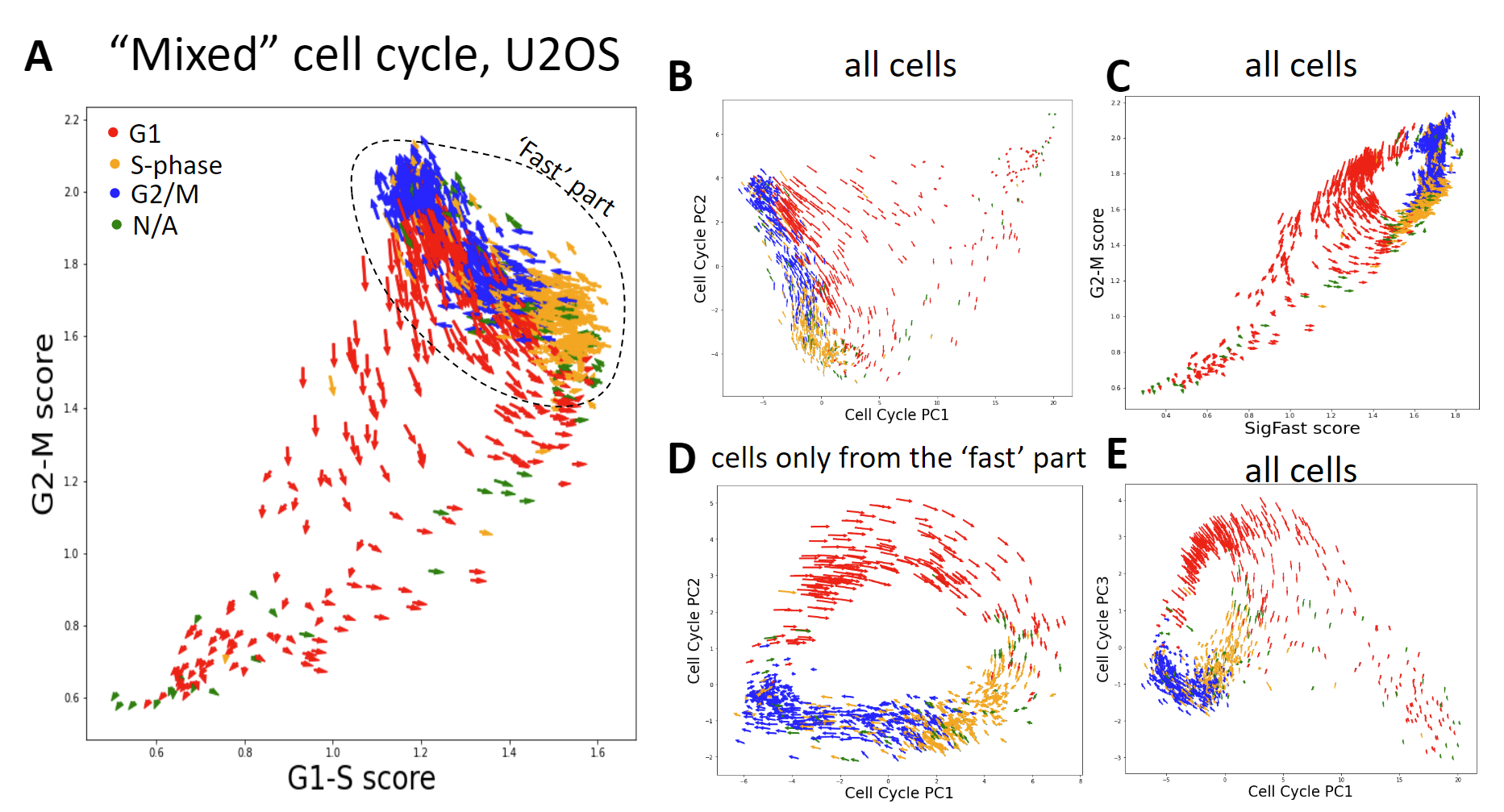} 
  \caption{Mixture of normal and "fast" cell cycles for U2OS cell line profiled using scRNA-Seq. The data points are colored by FUCCI phase labels.
  {\bf A.} G1/S-G2/M plot 
  {\bf B.} PCA plot from cell cycle genes of the Turosh's gene sets
  {\bf C.} The fast signature plot 
  {\bf D.} PCA computed taking into account only the "fast" part of the trajectory
  {\bf E.} PCA components 1,3 computd in the space of the cell cycle genes 
  }
  \label{fig:doubleCellCycle}
\end{figure}

{\bf Example of U2OS dataset. } Recently a scRNA-seq dataset for U2OS cell line was produced with reasonably many cells (1100+) \cite{Mahdessian2021}, with single cell RNA-seq profiles annotated accordingly to the cell cycle phase using FUCCI markers which is the golden standard used for this purpose.
Let us argue that both "standard" and "fast" subpopulations are present for this dataset.  


Visualizing the dataset using G1/S-G2/M plot Figure~\ref{fig:doubleCellCycle}A or  PCA projection from the space of the cell cycle Tirosh genes Figure~\ref{fig:doubleCellCycle}B  one can clearly see the "standard" ("triangle") cell cycle pattern. The second ("fast") cycle is not so clearly seen - it is shrinked with S-G2-M edge of
the triangle - the way "fast" cell cycle always appears on G1/S-G2/M plots (see previous subsection).
Indeed, one sees G1 subpopulation (red color), which is placed on that edge. 
To "unshrink" the "fast" cell cycle and more clearly see both ones   
there are several  ways:
1) Using CC\_SigFast vs G2/M phase plot - the good quality data can reveal the existence of two embedded loops (see Figure~\ref{fig:doubleCellCycle}C). 2) Another approach consists in removing the "normal"-G1 related segments of the dataset and visualizing the "fast" part through PCA projection in the space of cell cycle-related genes (such as the ones defined by Tirosh) - see Figure~\ref{fig:doubleCellCycle}D - one clearly sees the loop - the "fast" cycle  which was shrinked in G1-S/G2-M plot now becomes evident. (See also \ref{fig:Collection}G).
3) The third way -  to look on principal components of PCA projection from the data space formed by the same cycle-related genes. See Figure~\ref{fig:doubleCellCycle}E where two cycles can be seen. 
All the three ways allow to see two cycles paired by the common S-G2-M part and, thus, confirming the presence of the two subpopulations with normal and "fast" cell cycle patterns. 

One of the main difficulties analysing the cell cycle pattern where two types of trajectories co-exist within one cell type, seems to be the following: the two cycles might share a common segment S-G2-M, like "Siamese twins". The clear difference between two trajectories is manifested only in the G1 part. This is expected because the ESC-like cell cycle trajectory is expected to have a "shortened" G1 part, which we clearly see in scRNA-seq data. For the standard cell cycle G1 can be split into two segments (post-mitotic and pre-replication) as described above, while for the fast cell cycle the G1-associated segment leading to the S-phase has a much shorter length. Biologically this means that after the start of the S-phase main transcription programs are quite similar for cells in "fast" and "standard" modes.
The challenge is how to find some marker genes which would distinguish the cells which will be determined to follow the "fast"/"standard" routes after the mitosis. Literature suggests that it might be related to cyclins E and CDK2, but it is not yet clarified on the transcriptome level. 

Double cell cycle pattern has not been noticed either in the original paper, where G1-part of "standard" proliferating cells was neglected as being "an obvious exception (cluster 5 in Extended Data Fig. 7a)" \cite{Mahdessian2021}, neither in the recent cell cycle analysis package \cite{Tricycle} where oppositely, "fast" proliferating G1-part was neglected since the package was mainly developed keeping "standard" cell cycle pattern in mind.  

Double cell cycle pattern is not very often pattern, however it appears in number of examples.
In addition to U2OS described above, one can observe it for MCF10-2a from \cite{Jeffery2021} (see 
Figure~\ref{fig:Collection}G and scripts:
\href{https://www.kaggle.com/datasets/alexandervc/scrnaseq-mcf102a-p53-onoff-cenpa-overexpress}{https://www.kaggle.com/datasets/alexandervc/scrnaseq-mcf102a-p53-onoff-cenpa-overexpress}) -
that is in agreement with \cite{Spencer2013} where fast cycling cells were observed for MCF10 (however note that
MCF10-2a differs from MCF10, MCF10-2a it is highly chromosomally unstable and has closer to 80 chromosomes now). 
Another case is K562 from \cite{REPLOGLE2022} 
- see scripts:
\href{https://www.kaggle.com/datasets/alexandervc/scrnaseq-crisprperturbseqnormanselectedpart}{https://www.kaggle.com/datasets/alexandervc/scrnaseq-crisprperturbseqnormanselectedpart}.
We also suspect double cell cycle pattern for some breast cancer cell lines like BT549, CAL51, HCC1954, HCC114
from \cite{Gambardella2022} (scripts: 
\href{https://www.kaggle.com/datasets/alexandervc/scrnaseq-breast-cancer-cell-lines-atlas}
{https://www.kaggle.com/datasets/alexandervc/scrnaseq-breast-cancer-cell-lines-atlas})
that would be in partial agreement with \cite{Asghar2017} where subpopulations 
for breast cancer cell lines were analyzed and connected with CDK-4/6 inhibitors (like Palbociclib) resistance.
However, unfortunately,  at the moment our computational methods do not provide confident conclusions for these cases.


Our experience shows that as of today no package exists for the analysis of cell cycle trajectoriy from single cell data which would be capable of automatically treating the situation with the mixture of cell cycle trajectories. In the standard G1/S-G2/M plot the existence of the mixture is manifested as a triangular pattern with an unusually dense S-G2-M segment. 
Of course, all the difficulties described here are related to the situation when the two cell cycle patterns co-exist within one seemingly homogeneous cell population (same cell type or cell line). Otherwise, it can be advised first to separate cell types, and perform analysis on each cell type separately. 

Recent cell cycle studies (in particular \cite{Spencer2013} on mixture of the two cell cycles and role of CDK2)
has substantially improved understanding of the cell cycle regulation:
"Data from these ground-breaking studies have provided an entirely new perspective of the G1/S transition, and formed the foundation for a multitude of discoveries that complete the model" \cite{Hume2020Dianov}.
We believe that the use of scRNA-seq methods with adequate computational methodology for cell cycle trajectory analysis would promote further progress. 

\subsection{Summary} 
To conclude, we explained the interpretation of the G1/S-G2/M plot in this section..
In this plot, one can discover two main types  of the cell cycle trajectory: "standard" and "fast". 
Moreover there are unexpected cases when two cell cycles co-exist within seemingly homogeneous populations, e.g. one cell line. 
Most of the tools for the analysis of cell cycle from single cell transcriptomic data have been developed keeping in mind only one pattern of the cell cycle which creates certain difficulties in interpreting their results. Existence of two cell cycle loops embedded one into another represents even a stronger challenge for the existing cell cycle analysis methods.

\begin{figure}[t]
  \includegraphics[width=\textwidth]{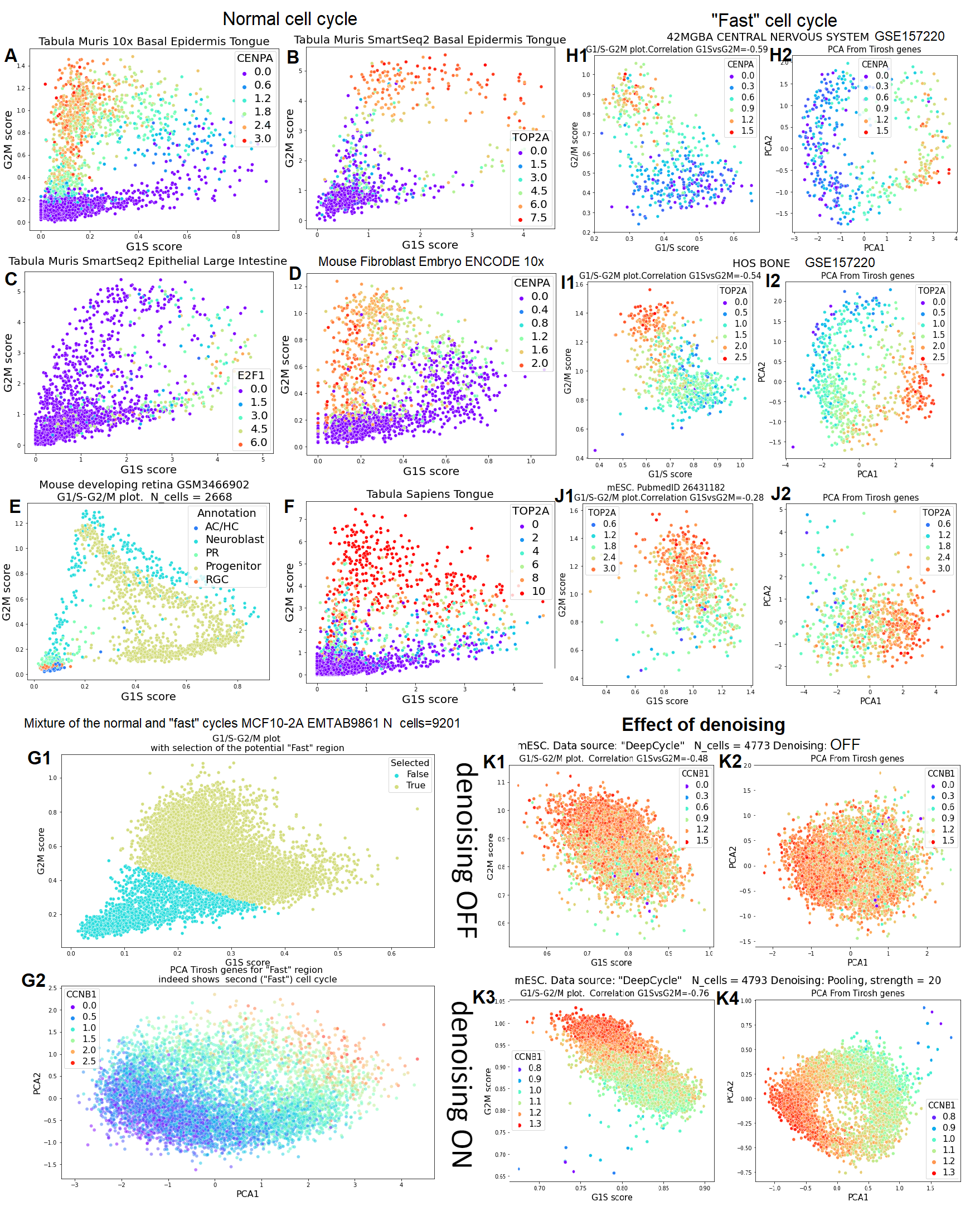} 
  \caption{Gallery of cell cycle trajectory analysis examples.
  {\bf A-F.} Normal cell cycle 
  {\bf G.} Mixture of the fast and normal cell cycles
  {\bf H-K.} "Fast" cell cycle; K1-K4 demonstrates that denoising is sometimes required to detect the "cyclic" pattern in the data point cloud
  }
  \label{fig:Collection}
\end{figure}
Some additional examples illustrating discussion above are given on Figure~\ref{fig:Collection}.

\begin{figure}[t]
  \includegraphics[width=\textwidth]{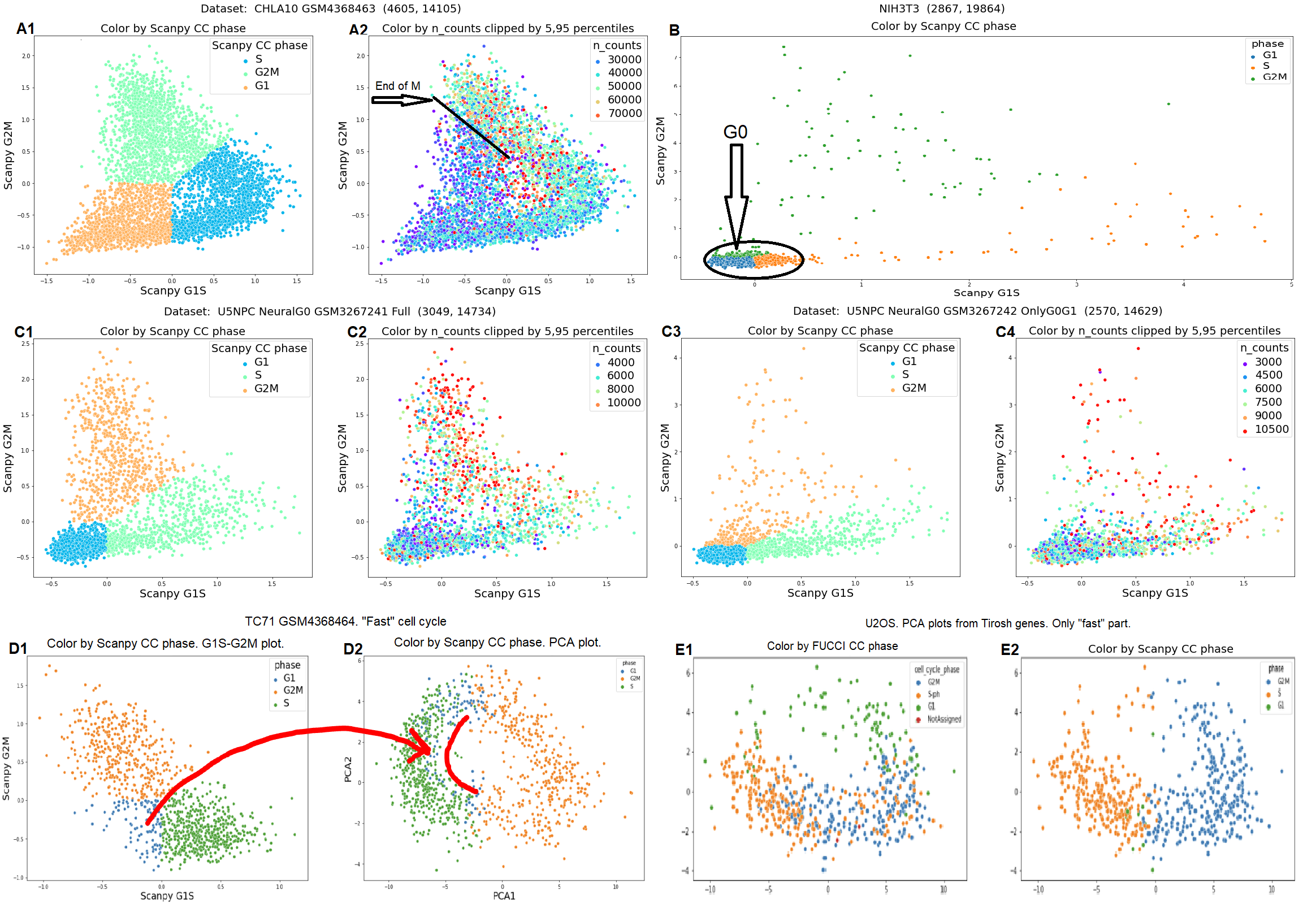} 
  \caption{Not quite correct cell cycle assignment by Scanpy.
  {\bf A.} Normal ("triangle") cell cycle pattern CHLA10 cell line \cite{miller2020Ewing}. A1 - Scanpy phases - G1 is underestimated.  A2 - compare with behaviour of counts.    
  {\bf B.} NIH3T3 cell line - most cells are non-proliferating - not consistent with Scanpy labels.
  {\bf C.} U5NPC cells \cite{CCAF} C1,C2 - full data; C3,C4 - filtered mainly G0-G1 - not consistent with Scanpy labels.   
  {\bf D.} TC71 cells  \cite{miller2020Ewing} - fast cell cycle - Scanpy G1 label has similar positions as S label along seen cycle shape on D2 plot.
  {\bf E.} U2OS - "fast" part only -  Scanpy labels do not correspond to FUCCI labels, G1 is absent - typical issue with Scanpy for "fast" cell cycle pattern.
  }
  \label{fig:SeuratScanpy}
\end{figure}
Let us also mention here that the baseline approach of Seurat/Scanpy not always work correctly. 
One may look at the figure~\ref{fig:SeuratScanpy} for pitfalls. 
More detailed discussion is given at subsections \ref{subsectCCphasesWhatPhases}
\ifbiorxiv
, \ref{sectSeuratScanpy}
\fi
.

\section{Five main computational tasks for the analysis of cell cycle progression from single cell data \label{sectTasks} }

The present section provides detailed discussion of the main tasks: descriptions, subtleties, approaches, state of the art, etc.

\subsection{Cell cycle (sub-)phase detection \label{sectCCphases} }



{\bf Task 1 (basic):} determine computationally cell cycle phase of cells from their transcriptome.


{\bf Task 1 (extented):} determine sub-phases with finer granularity: G0-state, G1-post mitotic, G1-prereplication (i.e. parts of G1 before/after Restriction checkpoint), subphases of mitosis, etc. 

\subsubsection{What cell cycle phases/subphases can be distinguished from single cell transcriptomes \label{subsectCCphasesWhatPhases} }

{\bf De facto standard Seurat/Scanpy methods to cell cycle phase labeling.} 
A simple and fast heuristic algorithm for labeling cells for their cell cycle phases based on the analysis of single cell transcriptomic profiles is implemented in Seurat (R-based) and Scanpy (Python) packages. Due to their popularity in the domain, the method can be considered as baseline for this particular task. The algorithm originates from one of the seminal papers introducing the single cell approach to transcriptome \cite{Tirosh2016}. It distinguishes the 'G1', 'S' and 'G2/M' labels and does not identify 'G0' cells. The algorithm seemed to be derived for the analysis of the "standard" cell cycle pattern. 

Our experience of exploiting this algorithm allows us to make the following conclusions. 
1) The algorithm provides basically correct, but not very precise 
results when most of the cells in the dataset are proliferating and follow the standard pattern of cell cycle. G1-part is typically underestimated. The border between S and G2M phases is subtle question for all packages.  
2) In the case when most of the cells are non-proliferating, Seurat/Scanpy tends to erroneously label significant part of the non-proliferating cells to ”S” and ”G2M” phases. 3) if the cell cycle pattern contains ESC-like part ("fast" or "mixed" type in the sense described above and typical for ESC, iPSC and many cancer cell lines), then "G1"-label is not correctly assigned for many cells.
See figure~\ref{fig:SeuratScanpy} for illustrations.

First packages developed to analyze the cell cycle from single cell transcriptomes could distinguish only three labels: G1,S, and G2M (i.e. union of G2 and M phases). G2 and M were merged in one label because of the following limitations: scRNA-seq datasets are  quite noisy and distinguishing some small cell subpolutions is not always reliable. Mitosis duration is short than the other cell cycle phases, thus mitotic cells represent a relatively small subpopulation.

More recent approach such as Tricycle \cite{Tricycle} using the ideas of Revelio \cite{Revelio}) could offer a finer granularity. For example, "G2M" labeling was splitted into  "G2" (early G2) and "G2M" (late G2 and M). Moreover they distinguished postmitotic G1 ( denoted "MG1"), and pre-replication G1 (denoted as "G1S" there). The analysis is based on 5 groups of genes which are more active in the corresponding cell cycle (sub)phases. These groups were proposed from a bulk transcriptomic study \cite{Whitfield2002}, and later updated/exploited in scRNA-Seq data in \cite{Macosko2015}. 

Determining mitosis subphases (prophase, prometaphase, metaphase, anaphase, telophase) is beyond the precision of existing scRNA-Seq technologies. Determining the G0-state is a special task which we discuss in a separate subsection.

\subsubsection{Basic approaches for annotating cells into cell cycle phases}

There are several strategies to assign the cell cycle phases to single cells. The first one is based on a careful selection of gene signatures, such that their behaviour (maximums, minimums, differences, etc) would allow determining the cell cycle phase. For example, as it is well-known from textbooks that the maximum of cyclin E indicates the end of the G1-phase and the S-phase start \cite{alberts2017molecular}. This approach is used in Seurat/Scanpy, reCAT,  Reveillo, Tricycle and DeepCycle packages. It is simple and interpretable, and our exprience suggests that it is also rather efficient and reproducible. 

The second approach consists in training a machine learning model on the datasets where the ground truth cell labels are available (e.g. FUCCI scheme) together with single cell transcriptomes. It is adapted in Cyclone, CCPE, cycleX, SC1CC, Pre-Phaser packages. The problems with the machine learning-based approach is that there are not so many high quality labeled datasets to train the models, and most of the available datasets are relatively old and small.

\begin{figure}[t]
  \includegraphics[width=\textwidth]{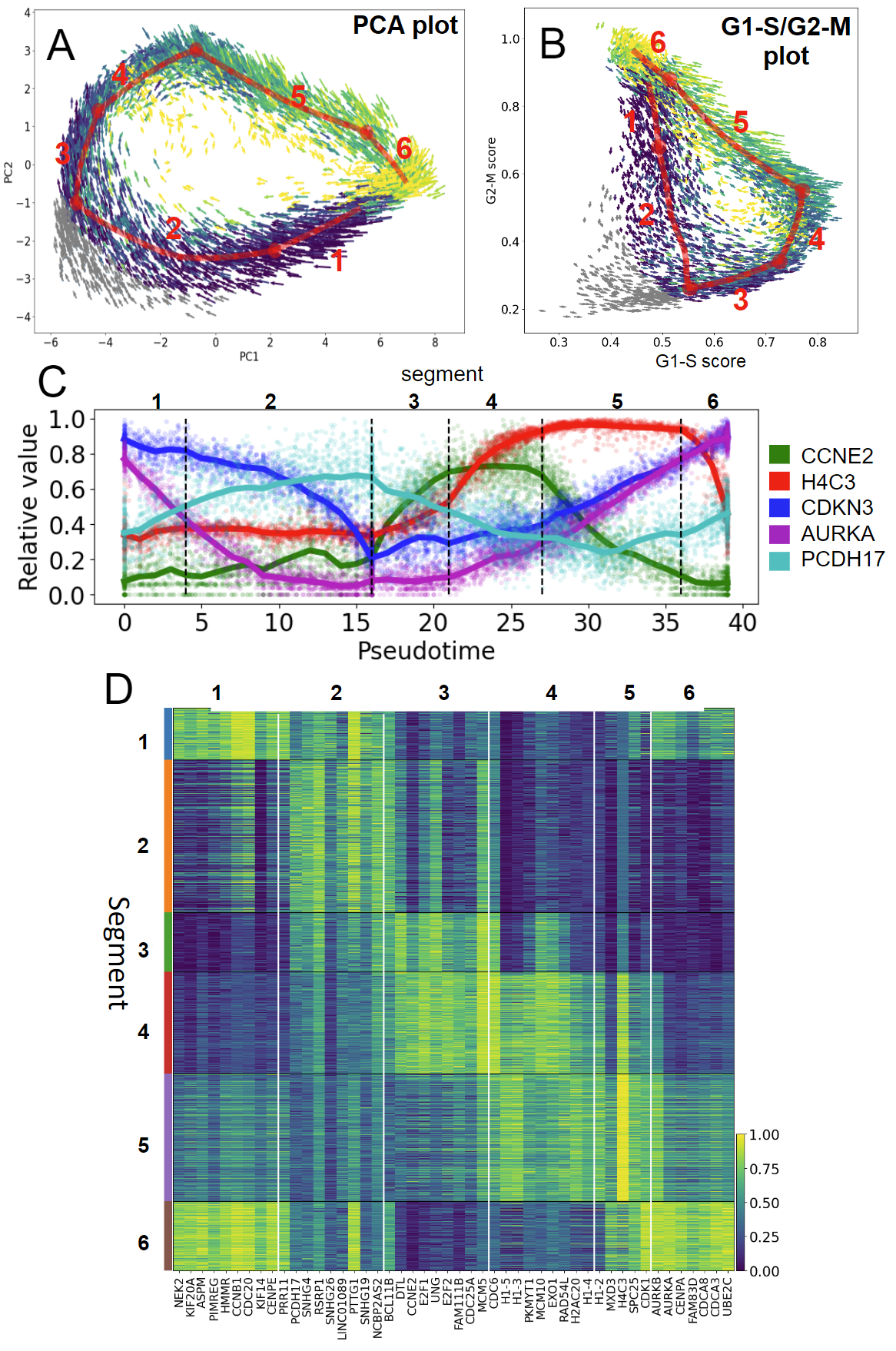} 
  \caption{The segmented structure of the cell cycle trajectory.  The  genes associated with each segment are shown using a heatmap. 
  {\bf A.} PCA plot. Segmentation of the the cell cycle trajectory by a piecewise-smooth approximation via ElPiGraph \cite{albergante2020robust}.
  {\bf B.} G1/S-G2/M plot. Same segments as in A) are shown.
  {\bf C.} Key representatives of the genes specifically expressed along each segment.  
  {\bf D.} Heatmap of gene (horizontal) expression vs. cells (vertical), for the cells arranged accordingly to the value of the cell cycle pseudotime.  
  }
  \label{fig:segmentedCellCycleV1}
\end{figure}

The third method consists in segmenting the cell cycle trajectory into relatively smooth fragments (see Figure~\ref{fig:segmentedCellCycleV1}A,B) such that each of the segments can be associated with a cell cycle phase. 
It also allows to cluster cell cycle genes to groups most active in each of the segments (Figure~\ref{fig:segmentedCellCycleV1}D).
The pseudotime profiles of selected genes from each group are shown on Figure~\ref{fig:segmentedCellCycleV1}C.
Such an approach was used in a recent study \cite{zinovyev2021modeling}. A general dynamic theory of self-dividing systems was suggested and a fundamental theorem connecting the number of segments $m$ in the cell cycle trajectory and its effective dimensionality $n$ was proven. The theorem states that $m\geq n$, and some considerations of general position type estimates that $m=n$. The approach is to be implemented as a computational tool.

\subsubsection{Existing packages and their quality of predictions}

The basic Seurat/Scanpy labeling into three cell cycle phases is rarely precise, but in a typical scenario is not hugely erroneous, see Figure~\ref{fig:SeuratScanpy}. There are three sources of errors: 1) for the datasets with mainly non-proliferating cells, some fraction of cells can be  erroneously  labeled as "S" and "G2M"; 2) in the case of the "standard" cell cycle the borders between phases can be set imprecisely, for example, the G1 labeled part can be smaller than the golden truth, 3) for the "fast" cell cycle pattern the G1 labels may be quite far from reality. 

Interesting experiment was performed in the PrePhaser paper \cite{PrePhaser} (Figure 1): run Seurat to get cell cycle phase labels, and then take only subparts of the dataset corresponding to each of the labels and run Seurat separately on each of the subparts. One would expect that the second run would return for each of the subdatasets only one label, but each subdataset was labeled in three sub-populations. The same issue is typical for many (with exceptions of Tricycle, PrePhaser) packages.  



Cyclone \cite{Cyclone} strongly tends not to put "S"-label (use only G1 and G2M) which is incorrect. The main point of the package reCAT \cite{reCAT} is to quantify the cell cycle pseudotime using TSP (Traveling Salesman Problem). It can also produce phase labels, which were reported to be adequate as in the original paper, as well as in Cyclum paper \cite{Cyclum} (Fig.2b), however these evaluations were made on relatively small, noisy and outdated datasets, while in more recent studies the results were less convincing \cite{Tricycle}. 
 It was also emphasized that "Note that although reCAT package provide function to assign cell cycle stage, it requires manual input cutoff for Bayes scores. It is unrealistic for us to pick some appropriate cutoffs for most of the datasets presented here." In practice this means that the choice of the parameters is not straightforward for reCAT package. Also reCAT performed poorly in evaluations in CCPE paper \cite{CCPE} (Fig. 3a,b). Certain problems of reCAT were also reported in SC1CC paper \cite{SC1CC} (Fig 3a)
 and in CCPE \cite{CCPE}  (Figure 2 and "reCAT ... do not characterize G2/M phase in the right order after S phase"). Overall,  use of reCAT for cell cycle phase labeling can not be currently recommended. 

The most recent among all packages considered here Tricycle \cite{Tricycle} (using some ideas from Revelio \cite{Revelio}) seems to produce the most convincing results. 
Nevertheless, it has a couple of drawbacks: first as authors write: "It is currently unknown what will happen if TriCycle is applied to a dataset without cycling cells". Second our own analysis shows that Tricycle may not always work correctly for the case of the "fast" cell cycle
\ifbiorxiv
(see subsection \ref{sectTricycle})
\fi
. 
Overall, for good quality scRNA-Seq datasets, containing the majority of proliferating cells Tricycle is worth to consider. 

For conceptual and also practical reasons, we do not recommend other packages (Cyclum, DeepCycle, CCPE, cycleX, SC1CC, Pre-Phaser) for performing the task. For example, Cyclum \cite{Cyclum} might have difficulties robustly   producing reasonable results on the datasets which are different from those on which it was trained. That is a conclusion from our own experience and also reported in \cite{DeepCycle}. The DeepCycle \cite{DeepCycle} is potentially very interesting, but it looks like more a proof of concept, rather than a ready to use tool.
Other packages ( scLVM, f-scLVM(Slalom), Oscope, ccRemover, Revelio,  Peco) do not assign cell cycle phase labels (or at least not in any straightforward way). 

\subsubsection{Conclusions on the  assigning cell cycle phases task}
Overall the task of assigning correct phase labels is not yet fully resolved. Nevertheless both basic approach of Seurat/Scanpy and the most recent TriCycle packages can be insightful for the analysis in many situations. 
From the first glance, the task does not seem very difficult, the basic idea is to use known gene markers of various phases. However, due to the noise in the single cell data, diversity of scRNA-Seq technologies which also evolve rapidly, the task is not yet universally solved. 

Certain problems for many packages arise for the situation when number of proliferating cells is close to zero. An experienced user should exclude that possibility during the preliminary investigation, since it is not yet done automatically by most of the packages.  

Many scRNA-Seq datasets contain cells of various types, and the cell cycle for them can be different. That creates an additional difficulty for the methods. Some methods require preliminary cell types separation, in particular, those based on the estimating the cell cycle "trajectory" like reCAT. 

The challenging situation is also the analysis of the cell cycle arrested under drugs or other perturbations. Examples of the dataset with hundreds of drugs tested with single cell readout are publicly available \cite{Srivatsan2020}. 

The idea of splitting cell cycle type to "standard" and "fast" must helpful for the cell cycle phase analysis, and it is not yet implemented in any approach reported here. 
Finally, we are lacking sufficient number of good quality datasets with FUCCI phase cell cycle phase assignment which can provide the ground truth for benchmarking this task. The only known to us good quality dataset with sufficient number of cells is U2OS \cite{Mahdessian2021} (see Figure~\ref{fig:doubleCellCycle}) but its analysis is complicated by the mixed nature of cell cycle pattern. There are several other scRNA datasets with FUCCI/FACS/Hoechst cell cycle phase labels, but most of them were obtained at the early stage of the scRNA-Seq era and so are quite noisy with small number of cells, so benchmarks using these datasets might be misleading. 

\subsection{Cell cycle (pseudo)time quantification}

Cells  continuously change their state during the progress through the cell cycle. The typical duration of the cell cycle for normal cells is around 24h (of note, there are reports on the coupling of the cell cycle  to the circadian cycle \cite{rougny2021FranckDelaunayCoupledCircadianAndCC} (and references therein), so 24h is probably not accidental). The main point about a typical scRNA-seq experiment is that it does not contain time labels for single cells, capturing a cell population as a single snapshot. So one arrives to the following task:

{\bf Task 2 (basic)} Quantify the degree of every cell progression from mitosis along the cell cycle by some continuous number (sometimes called a "pseudotime"). 

{\bf Task 2 (extended)} Relate the pseudotime and the actual physical time passed from mitosis.

The task formulated above is a particular case of the typical challenge for scRNA-Seq computational methods: to infer the temporal dynamics from a single cell population snapshot. A number of cellular trajectory inference" methods have been developed,
a recent benchmark study systematically compared 45 methods out of 70 \cite{saelens2019comparison}. However, they typically try to capture the pseudotime characterizing the cell differentiation process. Sometime these general methods are also applied for the analysis of the cell cycle, but using methods designed specifically for the cell cycle analysis are expected to be more efficient. 

\subsubsection{Brief overview of the cell cycle pseudotime quantification approaches}

In the next section we will provide a more detailed overview of cell cycle analysis packages, here let us only briefly give some recommendations. For some good quality and not complex datasets many approaches perform well and agree with each other. But, as for the other tasks, there is a single solution that would produce reasonable results for all the use cases, and typically there is no warning or confidence score which would notify about possibly wrong outcome. 


For example, TriCycle is fast and can work for datasets with many cell types and  produce reasonable results for datasets with the  "standard" cell cycle. The problem appears for the cases of the "fast" cell cycle and datasets with non-proliferating cells. The reCAT and PECO packages might  help for the cases of the "fast" cell cycle, at least when the dataset is not very noisy. 
Oscope was one of the first packages addressing this task. Unfortunately it is not fast enough to process datasets with thousands cells which appear routinely nowadays. DeepCycle is very innovative, but in the current state it is quite cubersome to use at least in the first line. 

Below we will also discuss such packages  as Cyclum, PrePhaser, CCPE, cycleX and SC1CC. 
The remaining ones such as Seurat/Scanpy, scLVM, f-scLVM(Slalom), Cyclone, ccRemover, Revelio do not provide functionality to estimate the cell cycle pseudotime (at least not directly). 





\subsubsection{Basic approaches for the cell cycle pseudotime quantification \label{subsectPseudotimeMethods} }

In order to solve this task, the scRNA-seq dataset is typically restricted to an appropriate subset of cell cycle-related genes. The rough idea is to identify a circle-like or oval-like pattern of the single cell data point cloud in this space. Then one just needs to parameterize the circle and it provides a pseudotime.



{\bf Embed to 2d and use polar angle.} A clever way is to embed the scRNA-seq dataset to 2 dimensions, such that the embedding should reveal the "oval" shape near the coordinate origin. After such an embedding is found, one can use the polar angle as a pseudotime measurement. This approach is adapted in Tricycle \cite{Tricycle} (and in some sense Revelio \cite{Revelio}).

{\bf Find a "circle" (cyclic graph) approximating the dataset.}  In this approach, no 2d embedding is necessary, the fit can be performed in all dimensions. This is a particular case of the cellular trajectory inference \cite{saelens2019comparison}. However, there is a specificity since we are searching for a very simple cyclic graph. Usually, a pre-selection of certain cell cycle genes is required.  This approach is used in reCAT \cite{reCAT} and ElPiGraph has functions to compute the cyclic principal graph \cite{zinovyev2021modeling,albergante2020robust,Gorban2008Principal,Gorban2007Topological}. The search for a cyclic principal graph can be in some applications considered as the classical traveling salesman problem for which many solvers are known: this idea used in Oscope \cite{Oscope}, reCAT \cite{reCAT}.

{\bf Machine learning approaches} can be used when the pseudotime is already known and the machine learning-based regression model is trained to predict it in new datasets. This approach is adapted in PECO \cite{Peco}, where the golden truth pseudotime is estimated from the FUCCI scores. 
Alternatively, one can build a neural network-based autoencoder with a one-dimentional bottlneck layer is assumed to have a circular "symmetry". This approach was used in Cyclum \cite{Cyclum}. A more sophisticated approach, with the use of the expression splitted into spliced/unspliced transcipts and the same idea of autoencoder was used in DeepCycle \cite{DeepCycle}. 




Let us discuss some possible subtle points in cell cycle pseudotime quantification.

First, many tools work incorrectly for the datasets that have a small number of proliferating cells.

Second, cell cycle trajectory might not represent a simple cycle. 
For example, for the cell cycle arrest in the G1 phase we should expect to see only the G1 part of the cycle. This case represents a problem for those approaches which are strongly based on searching the full circle in datasets, in particular those of reCAT, Cyclum, DeepCycle, etc. On the other hand, for the Tricycle, PECO, PrePhraser it is not a problem since they do not search for a circle. 


{\bf Execution time.} Tricycle and similar methods are computationally performant. Typically, the 2-dimensional embedding is computed by some linear projection which sometimes can be even precomputed. The trajectory-based methods like reCAT are less computationally performant even this does not usually preclude their use in practice. Training autoencoders can be relatively time consuming and even prohibiting in practice. 

{\bf Multiple cell types case.} Another advantage of Tricycle is that it can work for  datasets with a mixture of cell types. This situation can be a problem for many other methods in particular for trajectory based like reCAT. The reason is that despite the cell cycle is quite conservative and looks similar for many cell types (at least for normal cells), nevertheless there can be cell type-related specificities. The methods like Tricycle force projection to 2-dimensions thus mitigating possible differences caused by existence of multiple cell types, while the methods working in higher dimensional spaces are more sensitive to such differences.


{\bf 2d embedding is not always correct.}
Unfortunately, there exists no one single universal 2d embedding that would be suitable for all cell cycle trajectory types. In particular, the 2d embedding for the "fast" cell cycle seems to be different from the "standard" one. Probably, this can be resolved by using two embeddings, one for the standard cell cycle, another for the fast cell cycle.  On the other hand for the trajectory based methods it does not represent a problem as they search for the cyclic trajectory in the higher dimensional space not relying on any embedding. This consideration is similar for autoencoder based models. 

{\bf The pseudotime induced by polar angle is less precise than trajectory based pseudotime.}
The pseudotime induced by the trajectory methods has a very clear biological meaning, it directly reflects the amount of changes of cell transcriptome from one cell to another.  That is not quite true for the polar angle method because 1) a two-dimensional embedding distorts the dataset; 2) using the polar angle could be fine if the origin is placed in the center of the circle-like data point cloud. However, this might be difficult to achieve in practice: as a result, the dependence between the angular distance between two cells and the length of the hypothetical segment of cell cycle trajectory between them can be strongly non-uniform. 

As a separate argument towards the use of trajectory-based methods, it was shown that the length of the cell cycle trajectory is correlated to the actual physical time duration of the cell cycle (more precisely, cell line doubling time, see Figure 8 in \cite{zinovyev2021modeling}). Use of the polar angle as a pseudotime does not allow one to estimate the actual physical cell line doubling time.

\subsubsection{Controling the methods for quantifying cell cycle pseudotime}
There are several simple methods for evaluating the quality of pseudotime predictions
that allow user to be more (or less) confident about the obtained results. 

{\bf Agreement with visualization.} For good quality datasets one can often
represent the cell cycle organization using in 2 or 3-dimensional visualizations. For example, it can be achieved through the use of the G1/S-G2/M plot discussed above.
Thus coloring such a visualization by pseudotime - provides a simple and effective  visual inspection tool: one expects that the quantified pseutotime must correspond to the visually observed cyclic shape of the trajectory.  This works rather good in many cases, but not all. Subtle cases appear for noisy datasets (and is typical for the datasets produced long time ago), when the cyclic structure is not observed clearly.
Another even more subtle aspect consists in that the human visual inspection is limited to 2 or 3 dimensions, but the pseudotime  can be re-constructed in a higher dimensional space. Projection to 2 or 3 dimensions might bring some distortion, and it could be not clear either the pseudotime is constructed incorrectly, or one observes a meaningfull biological effect. 


{\bf Agreement with phases assignment.} The cell cycle order is G1,S,G2,M, so the pseudotime labels for cells should mostly agree with order of phase labels. For example, one expects that the pseudotime for all cells in G1 should be smaller than for all cells in S-phase, and so on. 

{\bf Expected pseudotemporal gene dynamics. } For certain genes, their expression profiles (e.g. maximums, minimums, etc)  along the cell cycle are expected to be consistent with respect to existing publications or databases such as Cyclebase \cite{Cyclebase3}.

{\bf Comparison with continuous FUCCI markerks.} FUCCI markers mCherry and EGFP allow one to give a continuous indicator of the cell cycle progress. Transcriptomic pseudotime is expected to be in agreement with this. The difficulty can be the lack of recent and good quality datasets with FUCCI labels. 

\subsubsection{Concluding remarks on cell cycle pseudotime quantification task}


Assigning pseudotime to cells is an important task for cell cycle analysis based on scRNA-seq data. Several tools are available but there are still issues to be resolved. One of the most recent Tricycle package offers the solution which should be tried as the first line. It is fast, robust, can work for datasets with multiple cell types,
but it might not always be fully correct, especially in situations of the "fast" cell cycle, "mixed" cell cycle type or in the case of datasets with too small number of proliferating cells. 

The task of relating the pseudotime to the real physical time is under-explored. Possibly it can be approached comparing known bulk studies (mentioned in section \ref{sectBulkTranscriptomicStudies} here) with scRNA-seq studies, as well as taking into accounts RNA-velocity and studying the systematic changes of cell density across various steps of cell cycle progression. Thus, larger density can indicate longer physical time spent by the cells at a particular step of cell cycle progression.   

\subsection{Detecting genes associated with the cell cycle progression, plotting their (pseudo)temporal profiles and clustering them}



Let us first give a general formulation of the task, and later split it into more concrete ones. 
It is quite related with previously discussed (pseudo)-time task, but is distinc from it in several important for applications aspects. 

{\bf Task 3 } Identify transcripts associated with the cell cycle via their expression pattern, quantify the association numerically,
compute (pseudo)time dependence of their expressions,
analyse the cell type dependence of these results. 

The genome-wide  studies (see section \ref{sectBulkTranscriptomicStudies} here) identified hundreds of cell cycle-related genes and provided their temporal profiles. However, despite these foundational results and multiple other studies, there is more work remains to be done. 

{\bf Task 3.1 } Identify transcripts involved in the cell cycle and   quantify the strength of their involvement. 

{\bf Subtask 3.1a } Issue: different studies report different lists of cell cycle-related genes. One needs to bring the order.

There are two reasons for this issue. First one is biological: for different cells (especially cancer cells) there exist differences in cell cycle regulation. The other reason is purely technical: measurements in different conditions and with use of various technologies
are characterized by different level of noise which affects the ability to distinguish cyclic genes. It is desirable to overcome that.  
In our experience from analysis of scRNA-seq data, the technical reasons play probably the major role in explaining the differences in explaining the different compositions of the cyclic gene lists. Few tens of the key genes which highly variable along the cell cycle are common for most of the studies but hundreds (if not thousands) with lesser amplitude appear in such lists not systematically. 
Therefore, it is desirable not only to report the binary fact that the gene is related to cell cycle or not, but also quantify
the strength of its association. That would allow us to compare the results across studies in more efficient way and probably distinguish technical and biological sources of variability.
Bringing the order to that situation is an appealing task - it is desirable to have some verified list of genes involved in the cell cycle, together with the quantified strength of the dependence and clear cell type specifications. The amounts of scRNA-seq data is growing every year by millions of cells that gives us hope that the technical difficulties can be overcome by using massive data.


Let us illustrate our point. Currently, the Gene Ontology term "cell cycle" contains  more than four thousands genes. Several years ago it was less than one thousand. It is rather clear that the involvement of most of  these genes in the cell cycle regulation is not central, and it would be quite useful to have some clear measure(s) of the "strength" of that involvement. Understanding how it is changing due to the cell type specificity is also important. The Cyclebase \cite{Cyclebase3} provides the certain measure of that "strength", scRNA-seq technologies may be used to improve it. 

Another example, the 4 main genome scale studies of the cell cycle dependent mamalian genes (discussed in section \ref{sectBulkTranscriptomicStudies} here) each reported several hundreds of genes but their intersection contained about one hundred. In our opinion it can not be explained by the biological reasons (since different cell types were considered in the studies). 

Let us mention that the package PECO \cite{Peco} provides functionality to estimate p-values for genes being truly cell cycle dependent, one the basis of permutation statistics. 


{\bf Subtask 3.1b } Not only protein coding genes are of interest, but also long non-coding RNA, micro RNA, alternative isoforms of some genes, etc. 

There are certain results that not only protein coding genes play a role in cell cycle regulation. For example, long non-coding RNA \cite{kitagawa2013lncRNA}, in particular MALAT1 \cite{Tripathi2013LongNRMALAT1} play a role in cell cycle regulation and related tumorigenesis. Frequently, MALAT1 appears in the analysis of cancer-related scRNA-seq datasets one of the most expressed transcripts.
It shows a significant pattern of dependence on the cell cycle pseudotime in several datasets. However, obtaining a fully consistent picture seems to be a task for the future studies. 

Other studies report on importance of micro RNA \cite{chen2010microrna}, \cite{gambardella2017impact}, \cite{deveale2022G1SmiR302} for cell cycle regulation. 
Inspiring paper \cite{dominguez2016high} writes: "we identified over 1 000 mRNAs, non-coding RNAs and pseudogenes with periodic expression".

Another aspect is paying attention to different transcripts 
which appear due to alternative splicing. Some results in that direction
has been reported in \cite{Mahdessian2021} (Extended Data Figure 7): it was observed that even for the genes most strongly related to cell cycle like BIRC5 and UBE2C there exist isoforms which cyclic dependence can be different. Moreover it was reported that "we found that a majority of cyclic genes (252 of 401) had both cyclic and non-cyclic transcript isoforms". For some genes like p73 it is widely studied that different isoforms play different roles in  cancers \cite{rozenberg2021p73}. 

Overall, the role of non protein coding transcripts and non standard isoforms in cell cycle is not well studied, and hopefully scRNA-seq technology might contribute to better understanding of this aspect. 

{\bf Task 3.2 } Quantify the dependence of transcript expression on the  cell cycle (pseudo-)time.


After the cell cycle pseudotime is estimated for individual cells, it is straightforward to quantify the (pseudo)time dependence for the transcripts. However, there exist several methods to perform this task, and due to the noisy character of the data the outcome might depend on the choice of the method and its parameters. Typically loess is used, but it might not be always the best choice. The reasons are the following: firstly, loess has a parameter, the sliding window length which has to be defined and secondly, is that loess might give some oversmoothing in some situations. 

For example, as we discussed above, the clearly seen configuration of the normal cell cycle trajectory in G1S/G2M plot has a triangle-like shape, so we might want to see these "angles" as clearly is possible, but smoothing can restrict that possibility. Unfortunately none of the packages pay enough attention to details of that step. In particular to deal with Task 3.1 one needs some statistical procedure(s) to distinguish pure noisy dependence from meaningful one which has not been implemented yet in any packages discussed here.


Producing (pseudo)temporal profile of proteins/transcripts is a classical tool. It can provide a clue to biological functions.  For yeast it was claimed that even 2-minute time resolution (about 1\% of the cell cycle time) could be achieved early in 2007  \cite{Rowicka2007}. Moreover, predictions of mutation effects was also achieved via application of mathematical modeling \cite{Chen2004Calzone}. Some recent bulk-RNA-seq studies reports a 20 minutes resolutions at least shortly after mitosis \cite{Chervova2022}. Of course, scRNA-seq approaches are quite far from such a precision, and even relation of pseudotime to real time is not explored, nevertheless these studies at least provide some goals to meet in the future.



{\bf Task 3.3 } Cluster transcripts into groups according to their (pseudo)time  dependence, analyse biological meaning of these groups. 

It is quite well-known that (pseudo)time profiles of many genes are similar. The genes with similar profiles can be combined into groups which often have biological meaning. For example, one transcription factor typically acts on a set of genes: therefore, these genes have similar (pseudo)time profiles. Well-known examples are E2F family of transcription factors which launch the G1S group of genes,
and FOXM1 transcription factor which activates G2M wave of transcription. 
Such a clustering of (pseudo)time profiles is typically done in many studies \cite{Chervova2022}: 
however, the existing tools typically do not provide such an opportunity as a ready to use solution. 

The most commonly used in scRNA-seq studies are the Tirosh G1/S and G2/M gene sets, which we already discussed. 
It seems G1/S, G2/M groups are quite fundamental - similar groups of genes can be seen for yeast, see Figure 4 in \cite{Rowicka2007},
moreover \cite{dominguez2016high} shows similar groups of genes can be seen as clusters in PPI networks, see Figure 4e in \cite{dominguez2016high}. 

The other widely used is 5 groups of cell cycle-related genes originating from \cite{Whitfield2002}, used in \cite{Macosko2015}, and recently revived in Revelio package \cite{Revelio}. 
Meta-analysis of bulk genome-wide studies performed in \cite{Giotti2017}, \cite{Giotti2019} offers extended cell cycle genes clusters. 
But these are some fixed gene sets. The task above is to provide computational methods for producing biologically meaningful gene clusters  based on particular dataset, and being able to compare them with existing references.


In order to conclude, the computation and analysis of cell cycle-related genes and their (pseudo)temporal profiling is classical and important task.
Although many packages offer construction of pseudo-time, and  thus computation of gene (pseudo)temporal profiles is straightforward, nevertheless the systematic analysis of such methods and results 
have not been performed yet. The desired analysis also includes: choice of cell cycle-related genes, certain smoothing for (pseudo)time profiles, and biologically meaningful clustering. These task have not yet deserved enough attention in the context of scRNA-seq studies. Hopefully this can be done and biologically interesting results can be obtained. 




\subsection{Distinguishing proliferating from resting cells, characterizing the "G0" heterogeneity \label{sectG0} }

It is well-known that most of the cells in mammalians are not proliferating.  There was long debate whether such cells are arrested in somewhere in the cell cycle program, or they are in some special state. Gradually the second hypothesis became prevailing.  Such a state is typically denoted "G0", or the terms "quiescence" and "senescence" are used. The "quiescence" means the cells which reversibly left the cell cycle,  while "senescence" means "irreversible" leaving the cell cycle (at least "irreversible" with our current level of knowledge).

The exact characterization of "quiescence" / "senescence" cells,   how cells enter/exit the cell cycle  is not yet fully understood. Such questions are quite important to cancer studies, because "cancer stem cells" assumed to be in some sort of "quiescence" state, while they are probably the reason why it is so difficult to treat the cancer. Other application: the tissue specific stem cells are also assumed to be in some sort of quiescence state, and only upon activation start to proliferate and differentiate to specific tissue cells. In that section we review certain computational  tasks for scRNA-seq analysis related to "G0" and the cell cycle.

\subsubsection{The "G0"-task(s): problem formulation}

{\bf Biological and technical aspects of the task.}
Actually there exist two related tasks. One is deep biological task - to understand the G0 state in general (how cells enter/maintain/exit "G0", what are biological markers etc). Another aspect is  technical and somewhat annoying. 
It consists in creating a robust computational algorithm for distinguishing non-proliferating cells from proliferating.
Most of the discussed here packages will produce incorrect results if the dataset would consist of non-proliferating cells (including the baseline Seurat/Scanpy). However, there is no warning or caution automatically generated about that situation even though the visual inspection typically can identify such situations. 

The general difficulty about these tasks is that there is probably no clear border between proliferating and non-proliferating cells. I.e. it seems that the textbook picture with some clear switch (crossroad), which splits the cells going to G0 from the cells going to the next cell cycle is not quite correct.  It is more likely that there is continuous intermediate area where cells may tend to quiescence or may tend to the next round the the cell cycle with some probabilities. It is well-known, that "quiescence" state is reversible "G0"-state, so cells may revert from that state and if that reversion happens quickly we simply would be able to catch the moment that the cell happened to be in that "quiescence" state, at least with modern scRNA-seq technologies which resolution is still limited. 
So it is more natural to search for some continuous characteristic which would estimate the "degree" of cell quiescence. 

Many packages typically fail on datasets with non-proliferating cells: one of the reasons is rather clear - packages like Oscope, reCAT, Cyclum, DeepCycle - are searching for a cycle/loop in the data cloud, but if cells are non proliferating the cycle does not exist, so one would produce just some random result. For other packages like Tricycle the reason is somewhat different, but, nevertheless, this situation represents a problem for it also.

Let us discuss the tasks in more details.

{\bf Task 4.1:} Develop/improve computational methods to distinguish proliferating  and non-proliferating cells.

Ideally one would like to have the ability to assign labels like "quiescence", "senescence" and continuous labels which would quantify how far the cell is from the proliferating state, since we expect there is not exact border between G1 and G0. However such kind of results might be difficult to achieve, and there is at least one case which we think is accessible and important to have for automatic cell cycle analysis packages: 

{\bf Task 4.1a (minimalist's technical task):} Be able to distinguish the situation when the whole dataset consists of mostly non-proliferating cells. 

Distinguishing such situation would at least save from incorrect search of the cycle structure in the data, where cycle does not exist. In our experience such situation is typically distinguishable at least through the visual inspection of the G1/S-G2/M plots. 
This approach which works rather well is quite well-known from the original works \cite{Tirosh2016}, \cite{Tirosh2016b},etc. The "G0" cells are  the cells where key cell cycle genes expressed on the low level. Indeed, on the G1/S-G2/M plot one can often see concentration of cells near zero - "G-zero near zero" - e.g. figure~\ref{fig:SeuratScanpy}B. However, the problem is to make a robust quantification of the "expressed on the low level". It stems to choice of certain thresholds which would mean "low enough", but the scRNA-seq data are diverse and choosing the thresholds which would robustly work across various technologies, cell types, etc, seems to be not fully resolved task yet. But hopefully  accessible.  The Seurat/Scanpy (based on \cite{Tirosh2016})  uses certain comparisons of the cell cycle genes with randomly chosen genes. This is a promising idea, but in our experience in many cases it does not produce satisfactory results.
In particular, it seems that Seurat/Scanpy systematically underestimate the "G1" part of the cell cycle trajectory.  For the datasets consisting of solely non-proliferating cells it labels them partly by "G2/M" and partly by "S", instead of labeling  all by "G1" (the "G0" label just does not exist at all in Seurat/Scanpy), see figure~\ref{fig:SeuratScanpy}B
\ifbiorxiv
and section \ref{sectSeuratScanpy}
\fi
.


More ambitious task consists in deeper understanding of the "G0"-state based on the scRNA-seq data.   

{\bf Task 4.2:} Quantification of the "G0"-state and its relation to the cell cycle.

{\bf Task 4.2a:} Quantify "degree" of "quiescence", i.e. produce a continuous label which would estimate how far is the cell from ability to proliferate, or, in the other words, how deeply the cell is in "G0". That seems to be more natural approach rather than searching for the exact border between proliferating and non-proliferating cells. 

{\bf Task 4.2b:} The current biological understanding that "G0"-state is not a single state, but there is continuum of different states, which arise in response to different biological stimuli, different kinds of stresses (starvation, presence of inhibitors, etc). Using the mathematical language, there exists a manifold of "G0"-states. The task is to define this manifold and relate its properties to the biological conditions.

{\bf Task 4.2c:} Define (improve/analyse already known) biomarkers for "quiescence"/"senescence", understand their dependence on the cell type and types of stresses imposed on cells (if any). 

{\bf Task 4.2d:} What are the possible transition trajectories  between the cell cycle and "G0"? (as discussed in \cite{Stallaert2022}, \cite{Stallaert2022PurvisPart2} there can be unconventional ones). 

We will review state of the art in the next subsection, but let us here mention the beautiful recent studies \cite{Stallaert2022}, \cite{Stallaert2022PurvisPart2}, which to some extent propose answers to almost all the tasks above and seem to be the most advanced contribution with respect to our discussion here. See in particular Figure~\ref{fig:FiveTasks}, Task 4. However, the papers were based on single cell proteomics (not transcriptomics), and  only 48 key proteins were measured.  It is interesting to try to reproduce (at least partly) these results with scRNA-seq approach, upon the success one may try to extend the results to various  cell types, taking to into account more than  a thousand scRNA-seq datasets are publicly available, which is not yet the case for single cell proteomics. 

Proteins dynamics on cell-cycle progression and cell-cycle exit has been explored in \cite{gookin2017Spencer}.


\subsubsection{Cell cycle and "G0": state of the art.}

Available packages provide tools to analyse "G0", we will describe them below. However, one may hope for more attention to this task and better results in the future. 

The original analysis in \cite{Tirosh2016}, \cite{Tirosh2016b}, etc. provided a simple and  clear way how simple visual inspection can approximately identify the non-proliferating fraction of cells 
("G-zero near zero"). Indeed, for many datasets one can clearly see a high density peak of cells in this region which corresponds to biological intuition that many cells are non proliferating in these datasets.
Such an analysis is enough for many purposes, but not for all. 
Besides above mentioned threshold problem, another issue is more conceptual. Using G1/S-G2/M plots as in \cite{Tirosh2016}, \cite{Tirosh2016b}, it is not always possible to distinguish those cells which are in cell cycle and just passing the "R"-point (the region transcriptomically nearby the true "G0"), and those cells which have left the cell cycle and are in true "G0"-state. The  expression of the key cell cycle genes e.g. G1/S-G2/M Tirosh is expected to be inversely correlated with the "degree" of quiescence.
However in our opinion, it is probably not the best measure for this purpose. Comparison of scRNA-Seq with the proteomics analysis in 
\cite{Stallaert2022}, \cite{Stallaert2022PurvisPart2}  one can hope to obtain more informative measures of quiescence.

In our opinion G1/S-G2/M plot of \cite{Tirosh2016}, \cite{Tirosh2016b} is an  invaluable tool for visual inspection, but more advanced algorithms are needed for automated treatment of the "G0 task".  Probably one should also use genes which are highly expressed in G0 and not only rely on low expression of the key cell cycle genes.

The package CCAF \cite{CCAF} is devoted to identification of the "G0"-like
cells in datasets related to neural cells.  The paper and the package are important contributions to the field
\ifbiorxiv
, see section \ref{sectCCAF} for detailed discussion
\fi
. 
The package is one of the rare packages which is able to put the "G0"-label. However there are several questions that are not that much clear for the moment: the focus on the paper is on neural cells - how far the classifier can be used/generalized to cells of the other type?
How it is related to analysis based on G1/S-G2/M plot? The paper produces a classifier - that means it puts precise border between G0 and G1, while it seems to be more natural to provide continuous label measuring how far the cell is in "G0", so it is it would be nice to understand better the biological meaning of the "G0"-labeled part.

The paper \cite{DeepCycle} where the DeepCycle package has been proposed also  contributes to studies of the "G0".  
\ifbiorxiv
We refer to section \ref{DeepCycleSect} for discussion of the package. 
\fi
In contrast to the previously mentioned paper the analysis is not achieved in fully automatic mode, but rather involves considerations specific to the question and a particular dataset. 

The package reCAT \cite{reCAT} has also an ability to put the "G0"-label.
However it seems the strong point of that package is to generate pseudotime, while assigning cell cycle labels may not be always correct and requires manually setting certain parameters. 
\ifbiorxiv
See section \ref{reCATSect} for detailed discussion
\fi
. 

The SC1CC \cite{SC1CC} is actually a web-site which can analyse the scRNA-seq datasets and cell cycle in particular. It also has the ability to generate the "G0" label. The approach is based on creating 7 clusters,
and  one would be assigned "G0" label. Example of "PMBC" dataset is correctly processed, i.e.  the dataset  consists mostly of non-proliferating cells and  cells were assigned expected G0-label . 
The package Prepraser \cite{PrePhaser} also has an ability to identify G0 cells. One needs to perform further benchmarks to clarify how robust these approaches are. 

Overall the situation with computational scRNA-seq analysis of the "G0"/cell cycle, is not as advanced as it might hopefully be. The recent single cell proteomics based papers \cite{Stallaert2022}, \cite{Stallaert2022PurvisPart2} provide some data to benchmark the existing methods.

\subsection{Removing the cell cycle effect from the single cell data}


{\bf The problem.} For cell populations containing proliferating cells, the cell cycle is typically quite a strong signal. So if one is interested in some other task: typically, identifying different cell types by clustering, then the transcriptomic heterogeneity related to cell cycle might typically be confounding. The reason for this is rather clear: clustering algorithms try to put into the same cluster those cells which have similar transcriptome, but the transcriptomes of cells in different cell cycle states are different (and sometimes that difference can even dominate the difference due to the cell type).

Similar issue exists for data dimensionality reduction and visualization methods. For example, in case of PCA one might be interested to see the difference due to cell type and so one might hope that the first principal components are related to the cell type signal, but often the cell cycle signal determines several first principal components. 

Thus the task to remove the cell cycle effect appears. 

{\bf Success story.} The first success of that idea was demonstrated by scLVM \cite{scLVM}, which showed that removing the cell cycle one is able to distinguish  biologically meaningful cell types in some datasets (see below). Since then more tools for that task have appeared. 

However, the task remains complex and probably less understood than the other tasks. Thus, each new paper typically shows quite substantial  problems for some previous approaches, but none offers an extensive benchmark on various datasets. Moreover, it is not always true that removing the cell cycle effect would help the clustering.


\subsubsection{Removing the cell cycle effect: problem formulation}

{\bf Task 5.} Develop robust algorithm(s) to remove/mitigate the effect of the cell cycle from gene expression.

Roughly speaking one wants to produce an expression matrix as if all the cells are frozen at the same position of the cell cycle. This would guarantee that the transcriptomic variance is not affected by the cell cycle or the influence of the cell cycle is minimized to some acceptable degree. 

Typically one needs to remove/mitigate the cell cycle signal, in order
to make other signals like cell type more distinguishable. Thus such a removal can be considered successful if it improves the clustering or other downstream tasks. The complete cell cycle removal is typically not the final objective by itself. The main  difficulty is that many genes contain variation from many sources, not only the cell cycle (see e.g.  see Fig 3a and Supplementary Figure 2  in \cite{scLVM}). There are examples where some packages over-correct, meaning they remove not only the cell cycle signal, but other signals too.  That is completely opposite to what is desired. For example, if we want to distinguish the cell types, but after cell cycle removal, the cell type signal is also degraded (see examples below) then such a correction can not be considered satisfactory. In practice, good quality datasets combined with state-of-the-art preprocessing+clustering algorithms might often achieve results even without cell cycle removal. So it seems to more dangerous to over-correct, rather than under-correct the cell cycle influence.  

{\bf Naive approach and its difficulties.} The naive solution to the task is very simple: just throw out (or assign constant values) all the genes known to be related to the cell cycle. However, there are too many genes which variation comes from both cell cycle and other sources. Such an analysis can be found in the first paper on topic - scLVM: see Figure 3a and Supplementary Figure 2  \cite{scLVM}. Thus freezing all the genes suspected to be related to the cell cycle will be over-correction,
but freezing only the key cell cycle genes will be under-correction because there many genes which are affected by cell cycle, but they are not key cell cycle genes. Therefore, solving this task is a subtle problem. 

The baseline approach of Seurat/Scanpy is regressing out from the total gene expression the G1/S and G2/M signals. That or another way all the methods conceptually stem fromt he similar ideas, but differs in the ways how one makes the predictions, in particular, using linear or non-linear methods. 

{\bf Removing the cell cycle effect always helps clustering ? No.} Sometimes removing cell cycle effect might be harmful for clustering the cell populations, for example non-proliferating cells and cells with the fast cell cycle would be clearly distinguishable by the cell cycle genes. 
So removing the cell cycle effect would decrease our ability to cluster populations. However, even in such cases, biologically one would be interested in some other differences between the cell types, rather than solely cell cycle genes, and search for such differences might be attenuated if the cell cycle is not removed. So there might be some subtle cases which should be treated not in the standard ways.


\subsubsection{Removing the cell cycle effect: state of the art}

There are quite a few packages which suggest methods for approaching this task: however, 
there is no extensive comparisons or benchmarks available in the literature. 
Careful development of benchmarks based on real, not simulated datasets and criteria to judge on the quality is a task on its own. 


{\bf scLVM.} The first paper and the package which proposed to reduce  cell cycle effect was a very well-known  scLVM \cite{scLVM}. The biological outcome was ”identification of otherwise undetectable subpopulations of cells that correspond to different stages during the differentiation of naive T-cells into T-helper 2 cells”.  The paper presented a package which can remove a confounding effect of any factor, the cell cycle is just a particular example. The input information for the algorithm contain
the list of genes which are mainly responsible for the confounding effect as an a priory knowledge.

{\bf ccRemover.} The next tool which appeared for this purpose is ccRemover \cite{ccRemover}. It extended the scLVM taking into acount the case 
when the main effects in the expression of the input list  cell-cycle genes may not be the  cell-cycle effect. The paper reports improvements over scLVM for several real datasets. However subsequent papers reported situations when ccRemover "over-corrected", i.e. it removes not only the cell-cycle signal, but the desired signal which distinguishes the cell types as described below.

{\bf  f-scLVM(Slalom)} \cite{fscLVM} is a follow-up package to scLVM. It extends its predecessor is several directions: it can work with several latent factors, and ”refines gene set annotations, and infers factors without annotation”. Moreover f-scLVM claimed to be scaled linear both with respect to number of cells and genes, in contrast to quadratic or cubic scaling of the other approaches.

\ifbiorxiv
More details on scLVM, f-scLVM(Slalom), ccRemover can be found in section \ref{FirstPackagesSect}.
\fi

{\bf Cyclum } \cite{Cyclum} is a profound neural network based (autoencoder) framework. It also allows a user to remove the cell cycle effect. The considered examples to remove the cell cycle contain one simulated data and one real data dataset, in both cases it was argued that Cyclum performs better than ccRemover and Seurat/Scanpy. 

Let us highlight the interesting original idea how to benchmark the cell cycle removing methods: the authors considered scRNA-seq melanoma data and analyzed two dormant drug resistance programs (MITF-high and AXL-high), which are known to be mutually exclusive by the previous biological analysis. However anticorrelation of the corresponding  scores
is not observed on the original dataset, and only after removing the cell cycle effect one gets clear anticorrelation between the corresponding  scores (see Supplementary Information Figure 7 \cite{Cyclum}).
This idea is quite in line with ideas typically considered in scRNA-seq denoising methods like MAGIC \cite{MAGIC2018}. 

{\bf Revelio} \cite{Revelio} is an interesting package developed in N. Rajewsky lab, with the focus of the study not on the package, but rather on underlying modelling  ideas, so there was no extensive benchmarks. It was argued that the cell cycle visualization / dimensionality reduction can be best achieved by linear methods (that perfectly agrees with our experience). In Revelio framework one removes the cell cycle effect rather easily just by subtracting the contribution of top dynamical components introduced in Revelio (see Figure 4B \cite{Revelio} and discussion around). 

{\bf CCPE} \cite{CCPE} is a more recent package. The example of the cell cycle removing considered here is quite interesting. One is advised to look at  Supplementary Figure S10 in \cite{CCPE} to appreciate the possible difficulties on the task, it is shown that ccRemover is over-correcting
such that we are unable to distinguish cell types, while Cyclum, Seurat allows one to see the clusters, but these clusters are not the ones which are biologically expected.  The dataset considered consists of 416 B cells. 
The CCPE package argued to work correctly on it, but it is the only real example which was considered in the paper.

{\bf SC1CC } \cite{SC1CC} is actually a web-site which allows to make the analysis of scRNA-seq datasets online, and in particular cell cycle analysis and removing the cell cycle effect.  The paper presents an interesting benchmark example of two distinct cell types  Jurkat  cells and 239T cells (Fig S1 \cite{SC1CC}). Despite cell types are different the ccRemover package over-corrects and mixes the cell types, that is inline with the example above. SC1CC performs well in certain examples, but again more benchmarks would be appreciated. 


To conclude, several promising approaches has been proposed to the task. A number of test cases has been worked out, showing certain successes, as well as issues. 
Applicability to situations of atlas scale datasets with many cells and cell types has not been explored. Moreover, unfortunately, in many cases newer papers present examples, where the previous packages not only under-perform, but perform incorrectly. Initial success of scLVM was based on the dataset with about 200 cells and two cell types. Nowadays the datasets are much larger in size, 
but on the other hand they may contain much more cell types (and this contrast makes the problem more complicated). Overall, a user may try to use the packages above: however, there can be some situations where they would not produce substantial improvements or even worsen the downstream analyses.

\section{Discussion. \label{sectDiscussion} } 

\subsection{Summary \label{sectSummary} }
Analysis of the cell cycle is an important  topic for scRNA-seq research
and the ability to analyse it on the level of single cells is 
crucial in many fields. Existence of many publicly available datasets open unprecedented opportunities to analyse the cell cycle for various organisms, cell types and perturbations in the uniform and systematic way. 

A number of packages has been developed specifically to analyse the cell cycle based on scRNA-seq data. However, as we tried to argue in the present review, there is more work to be done. The baseline package Seurat/Scanpy is fast, based on clear ideas and commonly used, but as we discussed above
it lacks precision as well as functionalities: it predicts cell cycle labels "G1,S,G2M" and remove the cell cycle effect, but there is no cell cycle pseudotime analysis, cell cycle-associated genes identification and profiling, "G0" labeling and analysis. There exist packages that offer the full spectrum of the approaches to all different tasks. However, as we discussed above, many issues are far from being fully resolved. In particular, there are many situations when packages may work incorrectly, but in most of the situations there would be no warning about potential problem, or confidence score returned which might alarm the user. This is general and important drawback of the available tools nowadays.  

On the other hand many issues seems to be quite approachable and hopefully would be resolved in some near future.

{\bf What is the "best" package.} The aim of the review is to emphasize that  none of the packages can be currently considered as the "ideal" one, because there are situations where each of them can produce erroneous results. So let us only make a very soft remark, that in our opinion the package which might be worth to try (after the baseline Seurat/Scanpy) and in certain situations is the most recent package "Tricycle" \cite{Tricycle}.  It is based on simple and  transparent  idea, it is fast, it can work on datasets with many cell types, the paper contains its careful tests on more than 10 datasets (in particular atlas style dataset) and comparison with many other packages. In produces cell cycle phase labels and estimates pseudotime. However, even it should be used with certain user's control: first as authors emphasize themselves it should not be used for the datasets with no (or almost no) proliferating cells, second for the datasets with ESC-like ("fast")  cell cycle we do not expect it to provide fully correct results, third the datasets with mixed cell cycle pattern like U2OS cell line, it does not reveal the presence of the two cell cycle trajectories.

So let us give some guidelines for users, which might be helpful to control and interpret the analysis, independently on what package is used.   

\subsection{Guidelines for users}

1. The first thing to do is to create G1/S-G2/M plot and try to roughly understand - does one have proliferating cells at all, how many of them, does one see "fast" cell cycle pattern. (See section \ref{sectG1SG2MplotFirst} ). In order to control what is going on, one may color the plot by the total number of read counts, expression of key cell cycle genes and cell types (if given). 

2. If one clearly sees the standard cell cycle pattern, meaning the triangular pattern shown in Figure~\ref{fig:StandardCellCycleWithExplanations} and, even better, the clear hole (region with low cell density) inside then the situation is not the most complicated one. One may further progress the analysis with a selected package (most of them will work well), and check that the produced results, cell cycle phase labels, pseudotime, gene profiles correspond to those  described in section \ref{InterpretationOfTriangle}. 

2.1. Sometimes one may see a triangular shape, but the "hole" is not clearly seen that typically means the substantial level of noise. One should be more careful in that case since some packages may  not work correctly. There are several tricks which sometimes improve situation:  a) denoising algorithms like simple read pooling, MAGIC \cite{MAGIC2018}, autoencoder denonoising \cite{Eraslan2019} (implemented in Scanpy)
 sometimes work well for that task and one can more clearly see the cycle with "hole" after denoising; b) one can try simply to increase filtering quality thresholds keep only the cells with less extreme number of total read counts, lower level of MT-score, etc. Sometimes it helps, but  there seems to be no universal solution, moreover dark side of denoising methods, that they may strongly distort the data if too strong denoising is imposed and that would lead to incorrect biological conclusions. 

3. If triangle pattern of normal cell cycle is not clearly seen then there are several possibilities. 

3.1. The fast cell cycle, i.e. the single segment visible from left-top to right-bottom at the G1/S-G2/M plot as shown in Figure~\ref{fig:review3typesOfCellCycle}. It is recommended to look at section \ref{sectFastCC}. One should be careful using Seurat/Scanpy and Tricycle in that case - most probably the results would not be correct. 

3.2. If one sees the concentrated point cloud near zero point in the G1/S-G2/M plot then most probably there are no proliferating cells and there is no point for the further cell cycle analysis. However, the subtle case can be when there is certain number of cells outside that cloud and these cells are suspected to be  proliferating. Sometimes it might be difficult to distinguish just outliers from truly proliferating cells. 
To some extent it depends on experience and on the number of cells: if the number of proliferating candidates is less than one hundred,
and they do not form clear cyclic shape, the interpretation needs to be careful. Such situation  would also be difficult  for most of the packages.
Some tricks which can be used are the following: one may try to separate proliferating candidates along with some number of non-proliferating cells,
and process that part only. In other words, one needs to avoid disbalance between proliferating and non-proliferating cells, which may cause
problems for many packages. Technically, in order to detect the proliferating cells, one can just impose condition: sum of G1S and G2M scores is greater than certain threshold, and the threshold is chosen such that it would drop out most of non-proliferating cells. 

3.3. If ones sees unusually dense right edge of the triangle, one may suspect that there is a double cell cycle pattern such as shown in Figure~\ref{fig:doubleCellCycle}. It is advised to read the section \ref{sectDoubleCC} and follow recommendations given there. This situation might be a difficult case especially if the level of noise is high. None of the scRNA-seq papers discussed this case and none of the  packages can automatically process it correctly to the best of our knowledge. 

4. In general, when analysing the cell cycle for a dataset with single cell type, we advise not to use UMAP/TSNE or other non-linear dimensional
reduction tools for the cell cycle analysis tasks. They  do not provide a standard interpretation, always produce different maps that would be impossible to directly compare with other datasete. Non-linear embeddings do not allow to visualize the linear segmented structure of the cell cycle trajectory, which is important since the segments might correspond to cell cycle phases, may lead to misleading conclusions, for example, ignoring the double cell cycle situation. We strongly recommend to use the G1/S-G2/M plot with Tirosh gene sets as the initial point of the analysis, but one can also use other linear methods.

To give an example of a distortions introduced by UMAP and erroneous conclusions made, one may look on Extended Data Figure 7a \cite{Mahdessian2021} cluster 5 (brown color)
and compare it with our Figure~\ref{fig:doubleCellCycle}A:  the cells outside the "fast part" marked region. These are the same cells. The linear plot shows that these cells occupy quite large transciptomic region, but UMAP distorted it into small cluster which was called "obvious exception",
while G1/S-G2/M identifies it as G1-part of the standard cell cycle . 

5. We do not advise to use too many cell cycle related genes to reveal the cell cycle structure. It is temping to think that the more cell cycle related genes one takes the more cell cycle information one gets, but it is not always the case.  Gene Ontology now annotates more than 4000 genes as related to cell cycle, but most of them are weakly related to the cell cycle and contain other biological signals as well. For example, making PCA from several hundreds of the cell cycle genes, may attenuate clear cell cycle structure seen from top 100 genes.  In our experience about 100 Tirosh genes work well in most of the situations, but probably one can find some improvements to that list.
So it is recommended  first to reveal the cell cycle structure with key cell cycle genes and only after that step start to analyse the rest of the genes using their pseudotime profiles. 

\subsection{Open  questions}
Development of computational tools probably would not be successful without better understanding of the underlying biological mechanisms related to the cell cycle. Let us mention some questions which in our mind would be important for both biological and computational sides, and several purely computational ones. 

1.  Molecular determinants of the switch between slow and fast cycles. In a sense, this question is about unglueing the "Siamese twins" the fast and the standard cell cycle trajectories (glued by their S-G2-M segments). 
This means that it would be desirable to find genes which expression differs (already in S-G2-M phases) the for cells which will be determined to go through the fast cell cycle trajectory, from those which would go by the standard cell cycle trajectory.  The literature suggests that CDK2 and Cyclin E should be involved \cite{Spencer2013}, \cite{Matson2017}, \cite{Asghar2017}, but it is not yet well understood on a transcriptome level. 

If the question would be resolved it would open new ways for computational treatment of the double cell cycle, which at the moment is not yet treated at all. 

2. How similar/different are cell cycles within one class (fast or standard) depending on cell types/perturbations.
Most of the cell cycles plots for the cells with the "standard cell cycle" looks quite similar (e.g., in G1/S-G2/M plot they form a triangle-like shape). However, looking at the atlas datasets like "Tabula Sapiens" \cite{tabula2022Sapiens} one can observe that nevertheless there exists small differences for some cell types. Can one describe these differences and understand the biological mechanisms which lead to them? The hope is to be able to do this using a small number of parameters, which knowledge would allow to characterize the time dependence of each of the thousands of cell cycle-associated genes. And one might hope for biological interpretation of these parameters. 

Speaking more mathematically each cell cycle is an embedding of the circle (time coordinate) to the thousand-dimensional space of genes expressions, but not all embeddings are allowed biologically, so the questions is to  parameterize the space of such embeddings and provide the biological interpretation of the parameters. Such general problem might be inaccessible, but as we see in examples, quite good approximation 
is provided by a triangle in G1/S-G2/M plot or more generally by a hexagon in PCA space of Tirosh genes. So the question about the differences of cell cycle regulation can be formulated at the level of the geometry of the cell cycle trajectory, measuring the segment lengths or angles of the polygon representing the trajectory and relating them to the context of the observation. Biologically speaking that is related to durations of cell cycle (sub)-phases. 

3. We briefly mentioned above that mutations of TP53 typically related to presence of the "fast" cell cycle. It would be interesting to understand effects of other mutations. In particular it might appear that mutations in retinoblastoma gene (RB1) would also lead to similar effect. RB1 belongs
to the same pathway as TP53, RB1 directly interacts with E2F1 (which is key transcription factor to launch the cell cycle). 

4. In this paper we discuss transcriptomic data, while classical works have been done for proteins, and the single-cell proteomics becomes more and more popular nowadays. However, the relation between transcriptome and  proteome measurements are not very well understood.  Shedding light in this direction would be quite desirable. Even the most basic question, labeling the cell cycle phases, is related to how proteomic measurements are related to the transcriptomic ones, For example, the classical textbook picture  shows that the maximum of Cyclin E corresponds to the border between G1 and S phases, and the same is assumed in scRNA-seq analysis. But this picture has been obtained for protein measurements, while currently is used for trasncriptome analysis, despite obviously it might not be fully correct.

5. Multi-omics data. Analysis of multi-omics is a major trend during the recent years.  The number of available multi-omics datasets is, of course, much less than scRNA-seq data, but hopefully it will grow in the future. It would be desirable to have computational methods ready for working with multi-omics data in truely integrative way. 

6. Nobel Prize for the cell cycle research was awarded in 2001, 
and in 2017 for another cycle -  circadian - for the "Discoveries of Molecular Mechanisms Controlling the Circadian Rhythm". The oscillation synchronization phenomena  is known since the 17th century. Nowadays, there is an active area of research which studies synchronization between the two cycles (e.g. \cite{rougny2021FranckDelaunayCoupledCircadianAndCC} and references therein). One may hope that scRNA-seq technique may also contribute
to this question. However, our own attempts were not quite successful yet, because the circadian genes are not highly expressed. 

7. The transition from S to G2 is not clearly seen neither on most of the visualizations, nor by the genes pseudotime profile analysis. It is quite in line with biological literature. The 2018 paper by Saldivar et.al. \cite{saldivar2018intrinsic} proposes a possible mechanism: "ATR represses FOXM1-dependent expression of a mitotic gene network until the S/G2 transition ensuring completion of DNA replication", while describing the prior situation:  "Much is known about the control of the G1/S, G2/M, and metaphase/anaphase transitions, but thus far, no control mechanism has been identified for the S/G2 transition". It would be desirable to improve our understanding of S to G2 transition: both from computational and biological points of view.

8. "Stem cell self-renewal is intrinsically associated with cell cycle control. However, the precise mechanisms coordinating cell fate choices and cell cycle remain to be fully uncovered " \cite{vallier2015cell}. 
One may hope that single cell RNA sequencing methods may pay a role in better understanding of that question. 
Some examples of the relation:
seminal paper by Pauklin and Vallier, 2013 \cite{pauklin2013Vallier} (brief review: \cite{dalton2013g1}) shows
that "hESCs in early G1 phase can only initiate differentiation into endoderm, 
whereas hESCs in late G1 were limited to neuroectoderm differentiation". 
Later on  Gonzales et al. 2015 \cite{gonzales2015deterministic}  (brief review: \cite{vallier2015cell})
showed that "factors controlling the G2/M phase are necessary to block pluripotency upon induction of differentiation".
It is also widely argued that short G1 for ESC is related to the fact that
differentiation happens in G1, and thus ESC keeps G1 short to avoid differentiation (see review \cite{neganova2008review}). 
It is interesting that on figure \ref{fig:Collection}E here for mouse developing retina one may indeed see that neuroblast cluster becomes distinguishable from the progenitor cluster after the end of G1, thus confirming the idea that differentiation happens in G1 . 
It also interesting to note that CDK2 plays a distinguished role both in the work of S.Spencer et.al. 2013 \cite{Spencer2013} 
(where fast cycling subpopulation was observed), as well as in the works of I.Neganova et.al. 2009,2011 \cite{neganova2009cdk2}, \cite{neganova2011cdk2}
(pluripotent status relies on CDK2 activity, knockdown of CDK2 resulted in hESCs arrest at G1 phase and differentiation to extraembryonic lineages;
downregulation of CDK2 triggers the G1 checkpoint through the activation of the ATM-CHK2-p53-p21 pathway; downregulation of CDK2 is able to induce sustained DNA damage...).
So better understanding how signaling, the cell cycle, and lineage specification are coordinated might hopefully be achieved with the help of single cell RNA sequencing studies.

9. Clarify list of examples of the cells which demonstrate the "fast"  of the cell cycles pattern.
As we discussed above the main examples for cells with the "fast" cell cycle pattern come from ESC, iPSC and cancer cells with broken TP53-pathway. 
However, there are probably more examples: 
remarks from \cite{Massague2004} suggests that "fast" cell cycle may be also observed for  angioblasts responding to vascular injury
and stem cells that replenishing the intestinal lining.
Recent remarkable paper by A.Singh et.al. 2022 \cite{Singh2022} strongly suggests that
"B lymphocytes undergoing multiple mitotic divisions, termed clonal expansion" should also demonstrate the "fast" cell cycle pattern
(however we were unable yet to find appropriate single cell dataset to check that hypothesis).
The remarkable properties of that process uncovered in \cite{Singh2022} (mitogen-independent proliferation, dependence on 
Survivin (BIRC5) to achieve mitogen-independent proliferation, etc.)  may probably be shared by the other cells with "fast"  cell cycle pattern, 
in particular by some cancer cells and ESC, iPSC cells. 
There are some weak hints in literature (e.g. \cite{CCAF}) that cells with broken Hippo/Yap pathway may also sometimes have "fast" cell cycle,
but that is far from being clear.
From our inspection of hundreds available datasets we do not expect much more examples of the cells with the "fast" cell cycle pattern.

10. Clarify the molecular mechanisms underlying the "fast" cell cycle pattern and possible mechanisms of switch from the normal to the "fast" cell cycle.
The literature and our analysis suggests that mechanisms are related to the p53-p21-RB-E2F-pathway. (See \cite{Engeland2022} (Fig.1) for a recent thorough discussion of the pathway). The mechanisms is roughly speaking the following: 
broken or suppressed p53 leads to lack of p21(CDKN1A) - which is direct target of p53 and is inhibitor of cyclin dependent kinases  - it mainly binds
to CyclinE/CDK2 complex, and thus un-inhibited  CyclinE/CDK2 phosphorilates Rb-protein, which allows  E2F transcription factor to start the hole program of 
the G1/S initiation. 
That is in line with huge amounts of facts:
our own analysis shows that "fast" cell cycle is related to TP53 mutation 
in cancer cells (Figure \ref{fig:cell_cycle_seesaw}C here);  
the pivotal role of CDK2 in the description above is in accordance with its role in S.Spencer et.al. 2013 \cite{Spencer2013} (fast cycling population is distinguished by CDK2), 
and I.Neganova et.al. 2009, 2011 \cite{neganova2009cdk2}, \cite{neganova2011cdk2} (crucial role of CDK2 for ESC cell cycle);   
for ESC it is well-known that
levels of TP53 are typically lower, moreover "critical role for P53 in Regulating the Cell Cycle of Ground State Embryonic Stem Cells" is described in \cite{Huurne2020}, in particular \cite{Suvorova2016} showed that "Nutlin, an antagonist of MDM2-mediated p53 degradation, restores p53-p21/Waf1-dependent G1 checkpoint in mESCs" that means restoring p53 activity for ESC one may achieve "normal", not "fast" cell cycle for ESC; 
it is also striking that p53 plays crucial role in iPSC production:
"blocking the p53 pathway vastly improves the ease and efficiency of transforming differentiated cells into induced pluripotent stem cells" \cite{Dolgin2009}, \cite{Kawamura2009}, \cite{Hong2009},  and so some authors  honored TP53 to be a "guardian of reprogramming" \cite{Menendez2010} in addition 
to the standard "guardian of the genome", see also "the p53–PUMA axis suppresses iPSC generation" \cite{Li2013} and 
recent review \cite{Fu2020} - that supports the idea that "shortened G1" (which is typical for the "fast" cell cycle pattern)  
is crucial for the pluripotent cells - both iPSC and ESC. 
Thus it is tempting to think that at least we have reasonable hypothesis on the mechanism underlying the "fast" cell cycle,
however further clarifications might be quite worth. 
In particular \cite{Engeland2022} mentions that p21 may affect not only CDK2 but other CDK also -  it is not clear how it fits the ideas above.
Recent discovery \cite{suski2022cdc7PiotrSicinski} (brief review \cite{kanemaki2022rethink}) changed
the understanding how CDK1 and CDC7 operate: it appears 
to be that in contrast to the previous models, CDK1 can be active
in G1 phase and substitute the role of the CDC7 and also affect the role of CDK2, it is not clear 
does that  new  discovery affects our understanding of the mechanisms underlying the "fast" cell cycle or not. 

Our own examples on the "mixture" of the fast and normal cell cycle within one cell population,
as well as previous results by \cite{Spencer2013}, \cite{Asghar2017} suggests that some cancer cells may spontaneously switch
from the "normal" to "fast" cell cycle pattern  and back without acquiring new mutations, and that phenomena might be related to the 
resistance to the CDK4/6 inhibitors like Palbociclib \cite{Asghar2017}. What are the mechanisms of that switch ? 
The paper \cite{Min2019} offers an idea that switch from the "fast" to "normal" cell cycle is a reaction of the cells to stress,
which allows them to go (if necessary) to the quiescence state  and survive stress in it. 
The proposed mechanisms are the p53 transcriptional response and the integrated stress response (ISR). 
However many questions remain unclear - analysing hundreds cell lines
we see that "mixture" of the cell cycles is rather rare phenomena,
what makes some cell lines distinguished in that respect ? 
What are the precise genes/pathways/etc involved ? 
For example, as we described above - downregulated p53-pathway is most probably the 
crucial reason for the existence of the "fast" cell cycle, 
can one give similar characterization of the 
cell lines demonstrating "mixture" of the cell cycles phenomena ?

And also lest mention several purely computational open questions.

11. Integrate RNA velocity ( \cite{la2018rna}, \cite{Bergen2021}, \cite{Kharchenko2021}) analysis with the cell cycle analysis.  RNA velocity shows  directions in which the transciptome will evolve in the nearby future.  Being restricted to the cell cycle genes it should be consistent with the cell cycle trajectory, and it is indeed the case as shown in many papers \cite{aynaud2020Ewing}, \cite{DeepCycle}, \cite{CCAF}, \cite{zinovyev2021modeling} etc. 

Thus RNA velocity currently is often used to provide additional justification that analysis is correct. One may hope even for more, it might be for some noisy datasets, RNA velocity combined with existing tools
would strengthen our abilities to work with noisy data. Anyway for the moment there is no package which allow automatically integrate RNA velocity analysis with the cell cycle analysis, but it is worth to develop one. 
One should also mention approach of DeepCycle \cite{DeepCycle} 
\ifbiorxiv
(see section \ref{DeepCycleSect} above) 
\fi
which  is very close to the idea of exploiting the RNA velocity signal, but it relies solely on the spliced/unspliced information, while the question here is to combine standard methods with RNA velocity ones. 

12. Actually preprocessing is one of the subtle sides of single cell RNA seq analysis, which might sometimes give substantial improvements for all down stream tasks and cell cycle analysis also.  The standard preprocessing is very simple: quality filtering, taking logs and library size normalization. But there are number of proposals how that can be improved. In particular, the standard library size normalization to constant size might be not the best for cell cycle analysis. Obviously the total number of read counts is expected to decrease by twofold after the mitosis. Doing the normalization
aimed to achieve the globally constant library size leads to loosing this phenomenon. More generally, one can observe that the total number of read counts is gradually increasing during the cell cycle progression which corresponds to the cell growth. There exists a promising idea of using the trajectory-based library size normalization, i.e. one may normalize counts 
to similar "constants" for cells which are close to each other along the cell cycle, but allow gradual variation of that "constant" along the cell cycle \cite{zinovyev2021modeling}.  The striking visually seen outcome of that procedure is clear break of the cell cycle trajectory after
the mitosis which corresponds to the fact that cells split into two parts but it is not seen with standard normalization. This normalization indeed can be done: see Figure 1 in \cite{zinovyev2021modeling}, but the issue is to make it work in "plug-in-play" mode robustly across various datasets, especially noisy ones. 

13. Important computational problem is a better understanding of what denoising methods (like MAGIC \cite{MAGIC2018}, Scanpy autoencoder \cite{Eraslan2019}, etc) are the most appropriate for  the cell cycle analysis. In our experience application of the denoising methods can  improve the analysis. Detection of the "cyclic" shape of the data point cloud often becomes substantially more clear: e.g. Figure~\ref{fig:Collection}K. However, too strong action of these algorithms leads to substantial distortion of the data and misleading biological conclusions, that is true in general and in particular case of the cell cycle analysis. What denoising algorithm and parameters are most suited for the cell cycle analysis is a widely unexplored question.


\subsection{Suggestions for package developers}

1. Nowadays, the sizes and the quality of the datasets are much higher than several years ago.  The older datasets with less than 1000 cells are probably not so good for benchmarking methods nowadays. In particular, those  with FUCCI and FACS cell cycle labels which sometimes considered as golden standards, may not optimal, because the level of noise in them is high
The recent U2OS dataset with FUCCI labels \cite{Mahdessian2021} is excellent and of high quality, but one should remember the subtle point due to presence of the double cell cycle pattern (see section \ref{sectDoubleCC} above). 

2. The methodology of G1/S-G2/M plots in many cases provides intuitively clear picture of the general properties of cell cycle trajectory, it is always worth to include comparison with it. 

3. It is worth to explicitly check the performance of tools on "standard" and  "fast" (ESC-like) cell cycle types,  since sometimes packages may work with datasets of one type, but not with the other one. 

4. The case of datasets with multiple cell types might be quite difficult case at least for pseudotime task, and the cell cycle removing task. 
It is worth to state explicitly: is the package intended to be used in such situations and then provide corresponding benchmarks. 

5. The datasets with no proliferating cells cause problems for many current packages.  It is worth to include the automatic detection of such cases and produce warning to users about potential problems.  

6. More generally, it is worth to include some control tools which would provide degree of confidence of the package in obtained results,
and inform user about potential issues. 

7. Large size of modern datasets impose constraints on performance, it is worth to provide some clearly stated time estimations  depending on the sizes of data. 

8. The situation with double cell cycle (see section \ref{sectDoubleCC} above) might be too difficult to process automatically, but at least it should be clearly stated as an exceptional situation, which should be analysed separately. The same about the cell cycle arrest, which can be found in datasets where drugs  are used as perturbations. Cell cycle arrest seems not being discussed in any of the packages, and  it seems to be a difficult case to analyse automatically, unless the control data without drug is given. 

9. The scRNA-seq datasets are quite diverse in terms of their biological context and  technologies, it is desirable to see benchmarks of tools, applied to as many datasets as possible. It our opinion having more than 1000 publicly  available  datasets, it would be worth to provide benchmarks on at least 10 or better 50 cases.

10. In general some ambitious computational goal can be creating packages which automatically process in "plug-n-play" mode large collections of datasets with hundreds examples, like the PANGLAODB benchmark corpus of data \cite{franzen2019panglaodb}, ARCHS4 \cite{lachmann2018archs4}. As an outcome one may hope to see the gene expressions pseudotime profiles for all datasets available. This would extend classical works like \cite{Whitfield2002} to a larger scale. Comparisons of the results for the different cell types might provide biological insights on how cell cycle depends on cell types and the biological context. 

\subsection{Use of Kaggle for analysis and the technical reproducibility}
The site www.kaggle.com is the very famous in Data Science community as a platform for data science-related competitions. Now it is subsidiary of Google and  directing to more general role "Your home for Data Science". 
It provides opportunity to store your data (up to 50G in one folder/dataset), store and execute code both Python and R (up to 9 hours continuous execution),  team collaboration, version control, share results with community and discuss projects, and all for free. 
From our point of view it is quite suitable for analysis of single cell RNA sequencing data and we frequently use it for our projects. Most of the datasets and code used in the review can be found there. 

The role of reproducibility is important in any research.  Kaggle is designed and polished for that by years of competitions and hundreds thousands users.  One needs to press just one button   - "Copy\&Edit" on any notebook to create your own copy  which will automatically have an access to the same data which original notebook has, and just running it - one can reproduce the results.  We strongly advise to use kaggle as a cloud service and sharing results with the community - one can keep private data, shared only within team and open it to community when the paper is published. There are more than 50 single cell RNA sequencing datasets already there (search on "scRNA-seq"), and  more generally hundreds bioinformatical datasets and notebooks,and about dozen of biologically inspired data science competitions. 

It might be interesting to organize the data science challenge on predicting the cell cycle phase labels and pseudotime. This would attract a large data science community to the cell cycle analysis and will push forward solving the tasks mentioned in this review.  For example, one challenge would be possible if unseen labelled (e.g. FUCCI) ground-truth datasets would be created and serve for evaluations of the proposed solutions (leaderboard).
One of the  co-authors of the important cell cycle paper  \cite{Mahdessian2021} has already lead two competitions on Kaggle \cite{kaggleHPA1}, \cite{kaggleHPA2}. So at least there is a positive experience of using this tool within the bioinformatics community.

\subsection{Conclusions }
Single cell RNA sequencing technologies continue to produce large amounts of data,  which will hopefully provide some important biological insights. The need in developing, maintaining, improving computational tools is of great importance.  Cell cycle is one the examples of the biological phenomenon (together with cell types identification, studying differentiation and other  dynamical biological processes) where single cell technologies have clear advantages. 

Moreover, the cell cycle is distinguished by the fact that any scRNA-seq dataset with proliferating cells can be used for its study: there is no need for special design of experiments, and thus the amount of data available is already enormous, but not yet thoroughly analysed.
Here we reviewed computational tasks arising in the cell cycle analysis,
and the computational tools to analyse it. 
Such tools started to appear around 2016 and now becoming more and more advanced. Still there are some issues to be resolved in the future. 

\section{Acknowledgements}

This work has been partially supported by the French government under management of Agence Nationale de la Recherche as part of the “Investissements d’Avenir” program, reference ANR-19-P3IA-0001 (PRAIRIE 3IA Institute) and by the European Union’s Horizon 2020 program (grant No. 826121, iPC project).




\bibliography{mybibfile}


\ifbiorxiv

\newpage
\appendix

\section{Technical review of the existing software features \label{sectPackages} }


The present section contains a brief overview of the discussed packages. 
Comprehensive benchmarks are out of the scope of the present paper.
The reader may look at the supplementary section of the Tricycle paper \cite{Tricycle} for the most comprehensive benchmarks available presently. 
The survey paper \cite{ml4scRNA} contains a brief section devoted to 
cell cycle analysis based scRNA-seq data with focus on machine learning 
approaches. 

\subsection{Cyclone, Oscope, ccRemover, scLVM,  f-scLVM (Slalom) \label{FirstPackagesSect}   } 
Cyclone \cite{Cyclone}, Oscope \cite{Oscope}, ccRemover \cite{ccRemover} are the first packages 
entirely devoted to the cell cycle analysis using scRNA-seq data.
They are devoted to single tasks: Cyclone - phase labeling, Oscope - pseudotime, ccRemover - removing the 
cell cycle signal from the data. 
At the time of writing this review, these packages seem to be outdated.
Thus, we noticed from experience that Cyclone can have problems with labeling the "S"-phase: in many examples it mainly annotated "G1" and "G2M" labels,
not recognizing the "S" cells;
Oscope is too slow to work on modern large datasets;
ccRemover reported to be over-correcting in some cases, i.e. removing too much variation from the data
which leads to loss of the biological information - in particular  cell type differences. 

Nevertheless, let us briefly mention some interesting findings from these papers, as well as scLVM and f-scLVM (Slalom) which are of more general use.

{\bf Cyclone \cite{Cyclone}. }
The package uses supervised machine learning strategy to assign cell cycle phase labels using the datasets with known labels for training.
From today perspective the datasets available at the time of package writing were small and noisy
(some having less than 100 cells).
Also the main training dataset was "182 mouse embryonic stem cells (mESCs)",
while ESC have specific "fast" cell cycle pattern it is different from normal cell cycle, so it potential source of the problem.
Overall it is not surprising that Cyclone benchmarks on the modern large datasets lead to not fully correct results - "S"-phase labels are almost missing. For example, see \cite{Tricycle} Supplementary materials Fig. S19 - one can see
clear "triangle" pattern of the normal cell cycle, but Cyclone marks cells by "G1" (green) and "G2M" (blue), the "S"-phase (red) is practically missing. This result appears incorrect to us, see subsection \ref{InterpretationOfTriangle} for explanations,
or compare with labels by the other packages in \cite{Tricycle} Supplementary materials). 
Nevertheless the methodology established for Cyclone development would probably be useful for the future development.  Six machine learning approaches were considered and the conclusions was that
PCA and pairs based methods were the most successful. 

{\bf Oscope \cite{Oscope}. }
The R-package Oscope estimates the pseudotime for the cell cycle. 
Unfortunately it scales poorly on large datasets. One of the probable explanations for this
is that it processes all pairs of relevant genes
("Oscope fits a 2-dimensional sinusoidal function to all gene pairs and chooses those with high scores" \cite{Oscope}),  
thus leading to "high computational burden" \cite{Cyclum} as was mentioned in many subsequent publications  \cite{Cyclum}, 
 \cite{Tricycle}.
The next steps of the Oscope algorithm: 
"Once candidate genes are identified, the K-medoids algorithm is applied to cluster genes into groups with similar frequencies, but possibly different phases. Then, for each group, Oscope recovers the cyclic order which places cells by their position within one cycle of the oscillatory process underlying the group." To define "cyclic order" is a kind of classical "traveling salesman problem (TSP)" (see also reCAT package discussion below).
The authors modify approach to the solution of TSP - "the nearest insertion algorithm":
"to order cells within an oscillatory gene group so that distance between each gene's expression and its gene-specific base cycle profile is minimized on average over all genes in the group", thus getting the cyclic order of cells as desired.
It is interesting that the approach can work not only on scRNA-seq data, but on bulk microarray data like the data from \cite{Whitfield2002}.
The examples from the subsequent papers 
e.g. Figure 3 \cite{Peco} show that Oscope produces reasonable results when the computational burden is not completely prohibitive.

{\bf ccRemover \cite{ccRemover}. }
The R-package ccRemover was created to modify the scRNA-seq dataset in order to exclude gene expression variance related to cell cycle.
It is one the tasks described in the main text of the present review and it is quite important for any downstream analysis of scRNA-seq data.  
Subsequent analysis in CCPE \cite{CCPE} (Supplementary Figure S10), 
SC1CC \cite{SC1CC} ( Fig.S1 ) demonstrates that ccRemover over-corrects the cell cycle signal,
that means it removes not only the cell cycle signal, but also the other biological signals,
such as cell type. The examples above show that different cell types become indistinguishable
after ccRemover for certain datasets. 
Some degradation of performance also reported in \cite{Cyclum} Figure 3 on simulated data.

The ccRemover was developed to relax some of the assumptions of scLVM, i.e. 
"the key assumption that scLVM makes is that all the main effects in the expression of cell-cycle genes are cell-cycle effects",
that means both packages rely on external list of annotated "cell-cycle genes",
but the key assumption of scLVM is that the main signal in these genes is due to cell cycle,
while ccRemover admits it might not be true. This is indeed seems natural since huge lists like modern Gene Ontology with more than 4000 genes related to cell cycle for sure contain not only the cell cycle signal and for many of these genes non cell cycle-related factors might be stronger. 
The analysis performed on several real datasets showed an improvement over scLVM, e.g. it was reported that scLVM improves by 2.5 percents classification quality over classification without cell cycle removing, while ccRemover improves 5 percents for "Real dataset 2: human glioblastomas data". 

The idea behind the ccRemover algorithm \cite{ccRemover} is "ccRemover carries out a simple PCA on the expression profiles of control genes to capture the sources of variation/effects...", and roughly speaking consists
in comparing the signal driving the control genes (not in the input cell cycle genes list) and the cell cycle genes (in the input list).
Those principal components where the cell cycle effect is greater are considered to be related to cell cycle. And  "then all effects declared as the cell-cycle effect are removed from the whole dataset by subtracting the projections of gene expression profiles on these effects." After that one repeats the procedure from the beginning "until no more principal components are identified as the cell-cycle effect". 

Overall the paper was an important contribution for removing the cell cycle effect task. 
The first tool addressing this task was scLVM \cite{scLVM} but which was of more general purpose  (not limited to removing the cell cycle effect)

{\bf scLVM \cite{scLVM}, f-scLVM(Slalom) \cite{fscLVM}. }
scLVM is a popular tool in the scRNA-seq field with almost 1000 citations up to date. It proposes some general framework to  remove the confounding effects,
in order to highlight the effect of interest.  The main example considered in the paper
is the cell cycle-related confounding effect for the objective of distinguishing the cell types. 
The  biological outcome was "identification of otherwise undetectable subpopulations of cells that correspond to different stages during the differentiation of naive T cells into T helper 2 cells". I.e. after removing the cell cycle effect the clustering into two cell subtypes became clear (Figure 3e - compare to Figure 3d (before removing) \cite{scLVM}).
One of the inputs for the algorithm is the list of genes which are suspected to be responsible for the confounding effect. 
The general idea of the methods removing confounding effects is predicting the values of variables
outside the list by the variables from the list; the variance which can be predicted is thought to be
due to the confounders and the difference between the value and prediction is the variance of interest.
The actual implementation is more sophisticated \cite{scLVM}, follow-up \cite{fscLVM} or the video recording \cite{StegleVideo}. 

The  f-scLVM(Slalom) is a follow-up package to scLVM, it extends the predecessor  is several directions: it can work with several latent factors, "refines gene set annotations, and infers factors without annotation".
Moreover, f-scLVM was claimed to scale linear both with respect to the number of cells and genes, in contrast to 
quadratic or cubic scaling of the other approaches. 
Overall scLVM and f-scLVM(Slalom) are important pioneering contributions to the single cell data analysis field, we give them less weight in the present review since they are not focused entirely on the analysis of cell cycle. On the other hand these works are widely known and referenced. 

\subsection{Seurat/Scanpy \label{sectSeuratScanpy}  }

Seurat \cite{Butler2018seuratV2} is the most popular R-based framework to work with single cell RNA sequencing data,
while Scanpy \cite{wolf2018scanpy} is its analogue for Python. There are tools for cell cycle analysis in Seurat \cite{SeuratCellCycle} which 
were re-implemented in Scanpy \cite{ScanpyCellCycle}.
They are based on seminal papers from  around 2015 by Tirosh, Kowalczyk, Regev et.al.: see Figure 2F in \cite{Kowalczyk2015}, Figure 2A in \cite{Tirosh2016}, Figure 3A in \cite{Tirosh2016b}, etc. which introduce G1/S-G2/M plots and corresponding analysis.
However, none of these publications provide a thorough evaluation of the approach, despite it is the most commonly used tool. 

{\bf Tasks.}
The tools are able to assign G1,S,G2M cell cycle phase labels and remove the cell effect, at the time of writing there is no pseudotime construction or the analysis of cyclic genes or
the analysis of G0 or labeling non-proliferating cells (such cells are labeled by G1).

{\bf Approach.} The labeling method is simple and is visualised in Figure~\ref{fig:SeuratScanpy}A1. The
region where both G1/S and G2/M scores are negative is labelled "G1". Otherwise, one compares the scores, and the region with G2/M greater than G1/S
is assigned G2M label, while the region with G2/M less than G1/S is assigned "S" label. 
It seems to be good enough approach for fast and very rough estimation, but is far from being precise in all cases. We believe that more advanced approaches can be beneficial for the use with modern high quality and large datasets.

The cell cycle removal algorithm consists of building the linear prediction model using a given input list of cell cycle genes as predictors,
and assigning the difference between the actual transcript values and predictions to be result of removal procedure.

{\bf Highlights and drawbacks.}
The advantages: simple, fast, robust. 
Disadvantages: the annotation is not always perfect, typically G1 fraction is underestimated.
The annotation is especially misleading in two cases. For the "fast" cell cycle pattern, G1 label can be absent at all or substantially smaller than the actual G1, see Figure~\ref{fig:SeuratScanpy}D,E.
For the case of the dataset containing mostly non-proliferating cells, a large part of them might be incorrectly assigned
to S and G2M phases, see Figure~\ref{fig:SeuratScanpy}B  .
For the case of the "normal" cell pattern Seurat/Scanpy precision might be also misleading with G1 fraction underestimated. However, it might be not so critical for applications where only a rough estimation is required.

Let us provide some details for the case of the normal cell cycle.
See Figure~\ref{fig:SeuratScanpy}A,C: A1 - Scanpy labels, A2 - shows the boundary of M and G1 which can be seen by the sudden drop of the total sum of the read counts which can be in used in high quality datasets for determining the moment of cell division.
One can see a substantial  difference of the border between M and G1 estimated by Scanpy (Figure~\ref{fig:SeuratScanpy}A1) and the one estimated based on the total sum of the read counts: G1 fraction appears much smaller than it should be.
The S phase label which is expected to appear at the right vertex of the triangle (see subsection \ref{sectG1SG2MplotFirst}), starts much earlier which probably leads to underestimating the G1 fraction. 
One can see similar effects in more noisy datasets in Figures~\ref{fig:SeuratScanpy}C1,C2. 
Figures~\ref{fig:SeuratScanpy}C3,C4 show the cell line data from \cite{CCAF}, but with preliminary filtering made in order to keep
mainly G0/G1 cells ("G0/G1 cells were first sorted on an FACSAria II (BD Biosciences) for the cells positive only for mCherry-hCDT1(aa30/120)").
Therefore, one expects to see mainly G1 label. However, Scanpy assigns substantial number of cells to be S and G2M. Thus G1 underestimation happens in that example also. The same phenomena can be seen in other examples. 

It is also worth noting the phenomenon described in  PrePhaser manuscript, in its Figure 1 \cite{PrePhaser}. If one takes a scRNA-seq dataset, computes the cell cycle phase labels, and then take only a subpart of the dataset corresponding
to G1 and again uses the tool only on that part of the dataset, the expected outcome should be to get G1 label for all the cells. But it is not the case for the Seurat/Scanpy methods.

{\bf Technical detail.} 
Let us mention that Seurat/Scanpy calculation of the G1/S-G2/M scores is not simply averaging of the appropriate groups of genes,
but involves an interesting idea to calculate the difference between these averages and  the ones calculated for  randomly chosen genes
with similar distribution (that should represent non cell cycle genes). Ideally that should give a way to determine the non-proliferating
cells. Since expressions of key cell cycle genes for non-proliferating cells should not be much different from the other genes
with the similar distribution. 
The idea is quite promising, but in our experience 
 it seems the implementation does not completely meet the desired objective. Also, z-score -based scaling is required by Seurat/Scanpy methodology which might amplify the noise/signal ratio. In our experience, we were not able to document a substantial advantage of using the Scanpy/Seurat-based gene set scoring compared to a simple averaging of gene expression in log scale, in the applications to cell cycle analysis.

To conclude: for the case of the normal cell cycle the vertices of "the cell cycle triangle"  (representing the "cores" of G1, S, G2M phases) are labeled typically correctly. However, the borders between phases can be wrongly annotated.
The annotation of cell cycle phases might be precise enough for some applications, and the method is fast and robust. The cell cycle phase annotation for the cases of the "fast" cell cycle and mostly "non-proliferating"  cells is not acceptable for most of the applications, since it might lead to incorrect biological conclusions like labeling non-proliferating cells as proliferating, and a huge underestimation of the G1 fraction. 

\subsection{reCAT \label{reCATSect}  }
The R-package reCAT (REcover Cycle Along Time) \cite{reCAT} is quite well-known in the field. One of the key ideas of it is the application of the travelling salesman problem (TSP) in order estimate the cell cycle pseudotime. Indeed it looks for the cycle in the cell neighbourhood graph. 

{\bf Approach.} The classical TSP searches for a cycle in a graph.  
The standard approach to get the graph from the dataset is to take a weighted graph representing the distance matrix or to exploit the k nearest neighbours (k-NN) graph. In both cases, in the case of single cell data, the graphs might be big and noisy. The standard trick to simplify the task if first to cluster the data 
and consider only the cluster centers instead of the initial data points. Afterwards one can consider the weighted graph representing the distances between the clusters centers and apply TSP. By aligning data points to the nearest graph nodes, on can quantify the pseudotime and obtain the cell cycle phase labels and the gene expression dynamics as a function of pseudotime.

{\bf Tasks.} reCAT provides estimation of the  pseudotime, cell cycle phase labels (including G0), identification of cell cycle-associated genes. It does not provide the cell cycle removing and distinguishing proliferative from non-proliferative cells.

{\bf Highlights and drawbacks.} The pseudotime estimation based on graph-based data representation and applying TSP is quite natural and beneficial compared to the polar angle-based pseudotime. 
Reports in the other packages like Tricycle ("The orderings
inferred by reCAT are largely consistent with our cell cycle position"), Peco ("we found that the orderings from reCAT agree most closely with the gating-based classification") confirm that in many cases the reCAT pseudotime estimation is reasonable. However the cell cycle phase estimation is probably not the strongest point of the package.  As reported in Tricycle it requires manual cutoffs, which are difficult or not always possible to set,
in SC1CC \cite{SC1CC} (Figure 3) a comparison is presented between the experimental cell cycle phase labels and the ones produced by reCAT. Also CCPE \cite{CCPE} reports certain issues 
(Figure 2 and "reCAT ... do not characterize G2/M phase in the right order after S phase.")

In our opinion the datasets which were used in reCAT paper were too small in size to correctly adjust the cell cycle phase labeling, as well as there were some other issues. 

To conclude: reCAT exploits a natural idea for estimation of cyclic pseudotime with TSP, which reported  to be mainly  consistent with other methods.  However, the cell cycle phase estimation is probably not the strongest feature of the package. 

\subsection{Cyclum  \label{CyclumSect} }
Cyclum Python package (using Keras with TensorFlow) is the first to use ideas of deep learning (more precisely autoencoders) 
to analyse the cell cycle using scRNA-seq data. 
Autoencoder is an unsupervised neural network which consists of the encoding part which typically compress the data
to get a reduced representation, and the decoding part which   regenerates the input from that reduced representation. 

{\bf Approach.} 
The application of an autoencoder appears a  natural idea to describe the periodic processes like cell cycle.
Indeed, training the autoencoder with the one-dimensional layer corresponding to the polar angle,  one may hope that such one-dimensional layer will provide a cyclic "pseudotime" for the cell cycle progression. 
Technically one considers not one-dimensional, but two-dimensional embedding, having  the form $(sin(x), cos(x))$, i.e. described by a single latent variable $x$. The Cyclum architecture is slightly more complicated: the bottleneck layer consists of the expected (sin(x), cos(x)) part which is extended by the linear variable ($X^{linear}$ - see the first equalities in Section "Methods" \cite{Cyclum}), so in a sense it tries to approximate the single cell data by helix. 

{\bf Tasks.} 
The Cyclum provides solutions to four main tasks of the cell cycle analysis: determining the cell cycle phase label,
calculating the pseudotime of the cell cycle progression, defining the cell cycle-associated genes,  and removing the cell cycle effect. The only task which is not considered in Cyclum is defining
the non-proliferating/G0 cells. 

{\bf Highlights and drawbacks.}
The Cyclum was the first package to apply deep learning to cell cycle analysis based on scRNA-seq data, which was an innovative approach. 
Cyclum can work as a out-of-the-box solution, provided with many examples of its usage and videotutorials \cite{LiangeVideo}. 
The package was applied not only to the cell cycle analysis, but also for studying EMT (Epithelial-Mesenchymal Transition).
It was compared with several methods available at that time and six real datasets were considered. The analysis by Cyclum of a relatively large hESC dataset with 12493 cell and 33694 genes takes about 40 minutes on a typical desktop computer.
The benchmarks made in CCPE \cite{CCPE} paper reported "the overall performance of Cyclum is better than Seurat, CYCLOPS and reCAT", but that results were reported on one of the datasets which was originally considered in Cyclum paper and is quite small. 
There are certain drawbacks, in particular benchmarks in DeepCycle paper \cite{DeepCycle},
reported poor performance of Cyclum on sevreal datasets. Our own tests lead us to similar conclusion. The computational time of Cyclum can be prohibitive compared to many other simpler methods such as Tricycle, which give result almost instantly. 
As for many other packages the case of non-proliferating cells is not discussed, and we can expect the results will not be adequate, since there is no warning about such possibility.
The datasets with many cell types were not considered. 
Overall despite strong and interesting idea of the approach, in our opinion the package cannot be recommended at least in the first line.

\subsection{Revelio \label{RevelioSect} }
Revelio \cite{Revelio} is an influential paper from N. Rajewsky Lab. The ideas presented there has been already used in other papers like 
Tricycle \cite{Tricycle}. The focus of the paper is more on theoretical modeling, with the package itself
supporting the paper conclusions,
so it does not seem to be an out-of-the-box solution. Nevertheless the ideas behind the package are interesting. 

{\bf Approach.} The main goal of the approach is to find  {\bf linear} a system of 2d coordinates called "dynamical components" (DC), such that the cell cycle-related data point cloud would look like a "circle" in these coordinates. To do it one restricts to the subset of genes related to cell cycle (using Gene Ontology and a cutoff for the expression); then one considers 3 dimensional PCA; and makes a change of coordinates in 3d, such that the
exluded region ("circle") would be seen in 2d while the third coordinate would correspond to a slower orthogonal dynamics.

{\bf Tasks.} The main focus of the paper is to prove that such construction is possible and can be effectively implemented.
The paper describes a method to estimate cell cycle phases (which was later improved in Tricycle), and emphasizes that the task of the cell cycle removal can be easily done due to the {\bf linear } nature of the method. Pseudotime is not discussed directly, but can also be computed from the Revelio approach by taking the polar angle in 2d space. 

{\bf Highlights and drawbacks.} One of the clear messages with which we would rather agree is that the {\bf linear} transormation of the initial cell coordinates is the most appropriate for analysing the cell cycle. In our experience use of non-linear methods: TSNE, UMAP, etc. are quite inappropriate to study the cell cycle
at least for the datasets containing one cell type. 
The cell cycle phase estimation in Revelio is at a finer granularity then in many other approaches: it uses 5 groups of genes and demonstrated that it gives reasonable results in several examples.  These 5 groups were found early in 2002 in \cite{Whitfield2002}, and later used in 2015 in \cite{Macosko2015} in scRNA-seq context, but were not widely used before the publication of the Revelio paper. 
 
One of the drawbacks is that some ideas described in the manuscript were not fully implemented in the package, but that was partially improved in Tricycle. The Tricycle paper provides several examples where Revelio approach does not seem to produce expected "circle"
\cite{Tricycle} ( Supplementary Materials Fig.S17). The reason is in the high heterogeneity of these datasets: in particular, they contain multiple cell types. In our experience this is expected and can be easily fixed. For example, one should not consider too many genes for computing the initial linear embedment: many genes from the Gene Ontology "cell cycle" contain not only the cell cycle signal. In our experience the Tirosh's set of genes sets exploited in Seurat/Scanpy is close to be optimal to find the correct representation because these genes' variation is almost exclusively related to the cell cycle.  
One can also at first calculate the pseudotime using a restricted genes set and after that check all the other genes. The dynamical (i.e. dataset dependent) nature of the coordinates has certain advantages, however, in our experience the fixed e.g. Tirosh G1/S-G2/M (as in Seurat/Scanpy) are more simple, more easy to interpret, allows to compare several datasets in the same system of coordinates.
It has an disadvantage not to show a "circle" for the case of the "fast" cell cycle datasets, but that can be fixed by considering a modified gene set, as we described above. 
Finally one of the papaer claims is that the shape of the cell cycle is a round circle. That seems to contradict to our experience as well as to the majority of visualizations obtained in other papers, and can be questioned from the known biological principles. 
Visualization of modern larger and good quality datasets typically show the "triangle"-like pattern (e.g. see subsection \ref{InterpretationOfTriangle}, or Tricycle \cite{Tricycle} ( Supplementary Materials Fig.S20). 
This matches better the model of existence of powerful transcription factors turning on and off, initiating the transcription of large groups of genes in one single moment, which is seen as turning points or segments in the cell cycle trajectory. For example E2F1 transcription factor initiates the G1 phase. The triangle seen in many visualization is certainly a simplification of reality,  nevertheless, the effect remain striking. 
We are convinced that the cell cycle trajectory in the gene expression space should be represented by a sequence of segments rather than round circle\cite{zinovyev2021modeling}. 

To conclude: the Revelio exploits interesting and innovative ideas,
but the package itself was not the main focus of the paper so it is not accomplished so far for its practical and universal use in cell cycle studies. 

\subsection{PECO \label{PecoSect}  }
The R-package PECO is a recent method to estimate pseudotime based on machine learning \cite{Peco}.  Previously machine learning methods predicted cell cycle phases, by training on datasets with  experimentally obtained cell cycle phase labels. The novelty of PECO approach is in suggesting a machine learning estimator for {\bf pseudotime} (not cell cycle phases), using experimentally obtained FUCCI intensities, which are continuously distributed. 

{\bf Approach.} The idea is to consider a dataset with both scRNA-seq and FUCCI markers available for a large set of cells. The authors generated their own dataset and then trained a machine learning regression model to predict the 
FUCCI intensities from scRNA-seq measurements. FUCCI intensities are continuous and one can calculate the cell cycle pseudotime from them.  

{\bf Tasks.} The main task resolved by PECO is cell cycle pseudotime reconstruction and it can also select cell cycle-associated genes. 
It does not provide automatic cell cycle removal and estimation of proliferative/non-proliferative/G0 cells. The paper discusses the prediction of the cell cycle phase labels based on partition around medoids (PAM) method,
but it seems to be restricted to the initial FUCCI-scores, not the predicted ones, so to the best of our understanding the package does not provide such functionality ready to be used in practice.   

{\bf Highlights and drawbacks.}
One of the interesting points of PECO was to demonstrate that only 5 genes (CDK1, UBE2C, TOP2A, H4C5, and H4C3) is enough to describe the cell cycle dynamics. Adding more genes did not improve  the quality substantially. 
Let us mention that the package PECO \cite{Peco} provides functionality to estimate p-values for  genes being truly cell cycle dependent, one the basis of permutation statistics.  The benchmarks in Tricycle paper \cite{Tricycle} 
( Supplementary Fig.S16 ) on 11 datasets report results consistent with Tricycle approach. In general machine learning methods are popular and powerful,
nevertheless we are not too optimistic about them for that task,
because scRNA-seq data quality is diverse and so the
models trained on some datasets may not generalize well
to the others. The number of datasets with FUCCI markers available is limited and they are of  quite small size, not representing the full diversity of the scRNA-seq data, many of these datasets are of poor signal quality. All that limits the applicability of the machine learning methods. As the authors of the PECO concluded by themselves: "Overall, our results suggest the need for more research and better data to quantify the accuracy and relative performance of the different available methods, including ours."

To conclude: PECO is a recent interesting method which 
seems to produce reasonable results as benchmarks mentioned above show.
However it seems to us other approaches e.g. Tricycle are sometimes
more robust for these tasks. 

\subsection{CCAF  \label{sectCCAF}}
The Tensorflow-powered Python package CCAF is a neural network-based classifier of the cell cycle phases with main focus on "Neural G0", i.e. it is one of the rare packages which can assign the "G0" label \cite{CCAF}. The focus of the paper is on the cells of neural origin, which is related to understanding the glioblastoma stem-like cells.
Cancer stem cells is an active field of research which might give improvements 
for treatment and understanding the cancer in general. Cancer stem cells are expected to be in some sort of "G0"-like state. 

{\bf Approach.} 
The authors first use unsupervised clustering to split cell cycle into several clusters, and then train neural network "ACTINN" (developed before \cite{ACTINN} in order to classify cell types) to identify these clusters in other datasets. 

{\bf Tasks.} 
Only the cell cycle phase labeling task is considered, and the focus is the identification of "G0" label.

{\bf Highlights and drawbacks.}
The paper itself is a deep study of quiescent-like cell state
in  "neuroepithelial-derived cell types during mammalian neurogenesis and in gliomas". The dataset produced is of excellent quality. The observation that  "knockout of genes associated with the Hippo/Yap and p53 pathways diminished Neural G0 in vitro, resulting in faster G1 transit" is in line
with the known general role of p53 and consistent with our own observations on the "fast" cell cycle (discussed above). Hippo/Yap pathway might give a clue to understand cases of the fast cell cycle not explained by p53. The package has been tested on non-neuroepithelial cells:  "actively dividing human embryonic kidney (HEK293T) cells" and reported to get reasonable results, as well as classifiers of  S and M phases by ccAF  were validated using the classical
synchronized bulk-microarray data \cite{Whitfield2002}. 
In general the structure of "GO" phase is not fully understood  (as we discussed above), and the ccAF paper provides interesting insights on that problem.  Recent papers from the Purvis' lab based on single cell proteomics, provided quantification and visualization of the degree of quiescence/senescence, thus not only identifying the "G0" cluster, but also characterizing its structure \cite{Stallaert2022}, \cite{Stallaert2022PurvisPart2}. One may hope combining the ideas of the two papers to shed more light on the organization of the G0-state in the future. As a drawback let us mention absence of comparison with G1/S-G2/M plots. The use of these standard plots would probably allow one to clearly positions the G0 subpopulation and also observe its structure as an elongated 2D part of the data point cloud near zero.
Thus it is not exactly clear how far the results are different from the standard approach, and how good the results of CCAF would be on the datasets with cells of other (non neural) origin having a substantial "G0"-part (in contrast to HEK293T which does not have G0 fraction), as well as other subtle questions on multiple cell type datasets or the datasets without proliferating cells.

To conclude:  the paper and package provide important insights. The package itself is not intended to be serve for general purpose cell cycle analysis, 
but rather has a specific focus, labeling the cell cycle phases in particular  "G0" for  neural and related cells. it is not yet fully clear more general applicability (other cell types) but that was not a goal of the paper.

\subsection{DeepCycle \label{DeepCycleSect} }
{\bf Approach.} DeepCycle  is one the most recent papers suggesting an approach to analyse the cell cycle using scRNA-Seq data \cite{DeepCycle}. It represents an innovative approach based on the use of the difference between spliced/unspliced transcripts to estimate the cell cycle pseudotime.
Thus, the idea is similar to RNA velocity approach \cite{la2018rna}, \cite{Bergen2021}, \cite{Kharchenko2021}. The approach is quite different from the other related approaches. On top of the use of RNAVelocity signal, a neural network-based autoencoder is used to combine the estimates for all genes into a single pseudotime measurement.

{\bf Tasks.} Deepcycle provides the estimation of the cell cycle pseudotime, cell cycle phase labeleing, identification of cell cycle-associated genes. It does not provide an automated cell cycle removing, estimating the proliferative/non-proliferative and G0-phase fractions. The paper contains a detailed discussion of the G0 part, but not in a fully automatic manner as discussed below. The obtained results for pseudotime and phase estimation  for three datasets considered look reasonable and  consistent with our conclusions. 

{\bf Highlights and drawbacks.}  The most innovative part of the approach is using the spliced/unspliced transcripts for the analysis of the cell cycle. What is impressive is that one of the considered datasets, mESC, contains large amount of noise (the cell cycle's "hole" is not seen by standard visualizations and that might be a problem for many approaches). Nevertheless, DeepCycle approach produces reasonable results in this case. This is a good indication that the approach is robust and potentially can work with not only the top quality datasets and be more robust than other approaches. Moreover, mESC cell cycle is not of the "standard" cell cycle pattern, but DeepCycle successfully deals with it. 

The cell cycle phase estimation is done after (and using) pseudotime computation, i.e. it is a separate function that takes the pseudotime estimation as an input and computes the cell cycle phases using the total read counts as another input. In our experience the total read counts is not always a reliable signal. 
It is not clear whether the approach would work well for the case of several cell types, and we see a potential problems here. Because it is not quite clear how consistently the number of spliced/unspliced transcripts depends on the cell type. The cell cycle phase estimation may also suffer in such situations. 
Moreover, the approach is based on a neural network autoencoder, an approach which might require fine-tuning before one can get reasonable results. So it is not clear whether the approach can always work in a fully automated way. In particular, the authors wrote that they excluded a part of the data before training for one dataset (the G0-like part), which is reasonable step to perform during the data preprocessing, but difficult to make it fully automated. 
The time to train autoencoder especially for large datasets (with tens of thousands of cells) was not estimated, but we can anticipate that it can be long. A technical, but important drawback is that for many openly available scRNA-seq datasets one can easily find the count matrices, but DeepCycle requires counts of spliced/unspliced transcripts which are much more rarely available pre-computed. If the splitted read counts are not available,  one has to perform such computations (that are quite heavy) using the raw sequence reads, and this also can be connected to difficulties (including legal ones) of accessing the data. 

To conclude: DeepCycle is an interesting innovative approach to cell cycle analysis: however, it might be too early to recommend it for the first line data treatment. 

\subsection{Tricycle \label{sectTricycle} }
A recent R package Tricycle is probably the package one may consider to try first today (after the baseline Seurat/Scanpy) \cite{Tricycle}. It uses straightforward principles, it is fast, robust and can work on datasets with multiple cell types. The paper offers one the most comprehensive benchmarks currently available (see "Supplementary Information , Additional file 1" \cite{Tricycle}) with more than 12 datasets considered, and compared with the previously developed packages:  PECO, Revelio, reCAT, Cyclone, Seurat.
Also "Tricycle is a locked-down prediction procedure. There are no tuning parameters" that is quite convenient for the user. Still there are certain drawbacks that we mention below.

{\bf Approach.} 
The idea of the method is surprisingly simple: one takes a high quality dataset, appropriate set of the cell cycle genes and computes the PCA embedding. After that, one re-uses that pre-trained PCA embedding for all the other datasets. The "transfer learning" name of the method is because it uses the transfer learning ideas from the modern machine learning: once pre-trained PCA embedding is transfered to the new data. 
It was shown for about dozen example datasets that this method produces an appropriate two-dimensional embedding of the data, such that it has "cyclic" shape around the origin of coordinates. Thus pseudotime is introduced just as a polar angle  on the plane. The cell cycle phases are typically assigned from the polar angle by the rule like this:  "approximately relate 0.5pi to be the start of S stage, pi to be the start of G2M stage, 1.5pi to be the middle of M stage, and 1.75pi-0.25pi to be G1/G0 stage".
Alternative cell cycle phase label estimation is also included in the  updated version of Revelio \cite{Revelio} 
based on the known key cell cycle genes which are top expressed in the corresponding phases of the cell cycle.

The biological reasoning for the  method to be successful is a strong conservation of the cell cycle across various cell types. This, however, has certain limitations as we discuss below. 

{\bf Tasks.} 
Tricycle provides an estimation of the  pseudotime, cell cycle phase labeling, analysis of the genes along the cell cycle. It does not provide automatic cell cycle signal removal, detecting the proliferative/non-proliferative and G0-phase fraction cells. 

{\bf Highlights and drawbacks.}
As we discussed above many times - Tricycle is probably the method to try, due to its simplicity, robustness and speed. Since its positive features has been already emphasized a lot, let us mention certain drawbacks.
First as authors mention by themselves - the package most probably produce incorrect results
on the datasets with no proliferating cells: "but we caution against using tricycle or other cell-cycle inference methods on a dataset without cycling cells."
Currently there is no internal control or a warning for a user, which  automatically generated for such situation.
The second issue is more subtle: it seems the Tricycle would not work correctly for the "fast" cell cycle pattern (subsection \ref{sectFastCC}), and moreover in case of the mixture of cell cycles (subsection \ref{sectDoubleCC}). The biological reason is quite clear: the Tricyle was trained on the example
of the normal cell cycle, while "fast" cell cycle is different and thus the transfer learning principle fails. 
More technically the reason is the following: first two principal components for the most of the datasets with normal cell cycle approximately aling with the standard G1/S , G2/M
signatures, but G1/S-G2/M visualization does not work correctly for the "fast" cell cycle pattern. 
To give a concrete example one may look on the Supplementary figure S20i \cite{Tricycle}, the example of U2OS cell line which contains both "fast" and "normal" cell cycle pattern (subsection \ref{sectDoubleCC}), while the figure shows recognizable "triangle" pattern of the normal cell cycle, and the "fast" part  is shrinked into single alongated segment, it is the same "shrinkage" as for G1/S-G2/M plot (see subsection \ref{sectFastCC}). The origin of coordinates is placed in such a way that one cannot assign the pseudotime correctly for the "fast" part. 
In our mind the idea to resolve that issue is quite simple: one should train two models, one for the "normal" cell cycle, and another one for the "fast" one. One can first determine the type of the cell cycle, "fast" or "normal", e.g. by computing the correlation between the G1/S and G2/M signatures scores, and then just apply the appropriate 
model. What is necessary to check how robust this approach will be.  The third drawback (of limited importance) we have already discussed in  subsection \ref{subsectPseudotimeMethods} is connected to the pseudotime inference task.
The pseudotime based on the polar angle is less biologically meaningful than the 
pseudotime computed as a geodesic distance along a trajectory, which one can potentially relate to the true physical time. 

To conclude: the Tricycle provides quite transparent, robust, fast, well benchmarked approach and it is worth to consider it. On the other hand as all the other packages it has certain limitations. 
Overall the power of the Tricycle is related to the biological conservation of the cell cycle trajectory, which allows us to transfer pre-trained embeddings to the new datasets.  On the other hand, different cell cycle pattern might lead to incorrect results. Hopefully, these limitations might be avoided in future using several pre-trained models and an appropriate model selection algorithm for choosing between them.

\subsection{ SC1CC, CCPE,  Pre-Phaser, CycleX \label{ SC1CCCCPEPrePhaserCycleX} }

There are several other packages for analysing cell cycle using single cell data such as SC1CC \cite{SC1CC}, CCPE \cite{CCPE},  Pre-Phaser \cite{PrePhaser}, CycleX \cite{cycleX}, that are probably less well-known. Most of these packages are not implemented in a standard Python or R framework, which probably restricts their use. Nevertheless, we will briefly review them here, highlighting some interesting points and ideas. 

{\bf SC1CC } (Single Cell One Cell Cycle analysis tool) provides a website
(https://sc1.engr.uconn.edu/ ) for the analysis of the single cell data in general and cell cycle in particular \cite{SC1CC}. The  methods rely on the "novel technique for ordering cells based on hierarchical clustering and optimal leaf ordering, and a new GSS metric based on serial correlation for assessing gene expression change smoothness along a reconstructed cell order as well as differentiating between cycling and non-cycling groups of cells".  More details: "Since the cell cycle is typically divided into 6 distinct phases (G1, G1/S, S, G2, G2/M, and M, see, e.g., \cite{cooper2007cell}), by default SC1CC attempts to extract up to 7 clusters from the hierarchical clustering dendogram corresponding to the 6 cell cycle phases plus at least one potential cluster of non-cycling cells – with a minimum cluster size threshold of 25 cells." 
And "Finally, to generate an order of cells consistent to their position along the cell cycle, SC1CC reorders the leaves of the hierarchical clustering dendogram (corresponding to the individual cells) by using the Optimal Leaf Ordering (OLO) algorithm".
The package has been tested on 7 datasets and compared with well-known packages ccRemover, Cyclone, reCAT. 
Benchmarking the cell cycle signal removal algorithm (see Fig.S1 \cite{SC1CC}) represents certain interest: 
one can see that ccRemover works not as expected (similar conclusion as in  CCPE paper - see below).
In general ccRemover seems to "overcorrect" the cell cycle signal. Figure 3 in \cite{SC1CC} shows the reCAT is not working as expected.  SC1CC is one the rare packages which is able to split proliferating and non-proliferating cells: the example of PMBC datasets demonstrates its correct work.  
As a kind of drawback  it seems that the implementation is not provided as standalone package, neither the source code of the web-site can be found on github. Also the standard G1/S-G2/M visualization is not provided, which is inconvenient, because it does not allow to visually control the quality of the obtained results.

{\bf CCPE} (cell cycle pseudotime estimation) \cite{CCPE} is a recent package using the idea is to approximate the data by a 3 dimensional helix with two coordinates proportional to sin and cos functions of an angle parameter, and the third one is a linear function of the angle (see equations 2 in \cite{CCPE}). A linear embedding from the higher dimensional space to fit the data is searched (equation 3 \cite{CCPE}). 
The package is tested on 3 simulated and 5 real datasets, and compared with well-known packages Cyclum, ccRemover, Seurat, reCAT.
It can estimate pseudotime, cell cycle phase labels, remove the cell cycle effect.
The benchmarking of the cell cycle signal removing (see Supplementary Figure S10 \cite{CCPE}) is interesting: one can see the difficulties of this task in case of application of Seurat, Cyclum and ccRemover that work incorrectly.
As a kind of a drawback let us mention its mixed implementation including Matlab, R and Python,
which probably restricts its usage. Also comparison with the standard G1/S-G2/M plot is not present.

{\bf Pre-Phaser} tool for the cell cycle analysis exploits a set of interesting ideas \cite{PrePhaser}. 
One of the points which probably should be kept in mind by everybody who is developing cell cycle scRNA-seq tools,
is the following: take an scRNA-seq dataset, compute cell cycle phase labels, then take only a subpart of the dataset corresponding to, for example, "G1" and again use the tool only on that part of the dataset. The expected outcome should be to get "G1" label for all the cells. But it is not the case for baseline Seurat/Scanpy, see Figure 1 in \cite{PrePhaser} . Actually, it is a typical problem for many cell cycle analysis packages. 
One could argue that taking a subpart of a dataset is an artificial situation, but it might play a role in the analysis of the cell cycle arrest (e.g., in "G1"-phase), and an ideal method should be able to deal with it. 
Of note, such an issue is not a problem for the "transfer learning" approaches like Tricycle \cite{Tricycle}.
The Pre-Phaser suggests that we can take into account also bulk microarray studies like \cite{Grant2013}, \cite{Whitfield2002}, etc. which are provided with timestamps, and the cell cycle phase labels.
The idea of the Pre-Phaser is the following: "we can compare its profile to the best-matching microarray (or other assay such as RNA-seq) time course data series points, to find the precise angle position. Since we want to use several best-matching time points to zero into the precise position, we employ nearest neighbor methods".
That is an interesting idea to compare modern scRNA-seq studies with the seminal ones
\cite{Whitfield2002}, \cite{BarJoseph2008}, \cite{Grant2013}, \cite{PenaDiaz2013}, which
have the actual physical time-stamps by their construction. So the comparison might give a clue
to relate "pseudotime" with the actual physical cell cycle time.  As a kind of drawback  let us mention its mixed implementation in C++ and Python, not availability of an easy and standard intstallation procedure. 

{\bf CycleX} \cite{cycleX} uses  Gaussian Process Latent Variable Model GPLVM as a computational core for the pseudotime estimation. It is implemented in R, with a small Python fragment.
Only one dataset with only 217 cells was considered. Cell cycle phase labels were estimated by the two methods: first by Cyclone \cite{Cyclone}, and the second with the help of the gene sets specific to five cell cycle phases  (G1/S, S, G2/M, M, M/G1) taken from \cite{Whitfield2002} and refined by Macosko (2015) \cite{Macosko2015}. 

\fi

\end{document}